\documentclass[11pt,a4paper]{article}
\usepackage{jheppub}
\usepackage{mathtools}
\usepackage{physics}
\usepackage{tikz-cd}

\newcommand{\orcid}[1]{\href{https://orcid.org/#1}{\includegraphics[width=8pt]{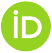}}}

\usepackage{natbib}
\newcommand{\pa}{\mathop{}\!\partial}

\newcommand{\R}{\mathbb{R}}

\def\be#1\ee{\begin{align}#1\end{align}}

\def\heq{\,\hat{=}\,}

\newcommand\nueq{\stackrel{\mathclap{\normalfont\tiny\mbox{$\cN$}}}{=}}

\newcommand{\diff}{\mathfrak{diff}}

\newcommand{\mr}{\mathring}
\newcommand{\fr}{\mathfrak}
\newcommand{\bd}{\boldsymbol}

\newcommand{\ie}{\emph{i.e.} }

\newcommand{\dext}{\text{d}}

\newcommand{\cS}{\mathcal{S}}
\newcommand{\cM}{\mathcal{M}}
\newcommand{\cN}{\mathcal{N}}
\newcommand{\cC}{\mathcal{C}}

\newcommand{\cF}{\mathcal{F}}

\newcommand{\cO}{\mathcal{O}}
\newcommand{\Lie}{\mathcal{L}}
\newcommand{\cW}{\mathcal{W}}
\newcommand{\cE}{\mathcal{E}}

\newcommand{\cK}{\mathcal{K}}
\newcommand{\cQ}{\mathcal{Q}}

\newcommand{\cP}{\mathcal{P}}

\newcommand{\cR}{\mathcal{R}}
\newcommand{\cB}{\mathcal{B}}
\newcommand{\cI}{\mathcal{I}}

\newcommand{\eps}{{\bd{\epsilon}}}

\newcommand{\f}{\frac}

\def\EH{\mathrm{EH}}
\def\EM{\mathrm{EM}}
\def\M{\mathrm{M}}
\def\EC{\mathrm{EC}}
\def\ECH{\mathrm{ECH}}
\def\H{\mathrm{H}}

\newcommand{\mtn}{\mr{\theta}^{(n)}}
\newcommand{\mpi}{\mr{\pi}}

\newcommand{\mKnu}{\mr{K}_{(n)}^{ab}}

\newcommand{\mthell}{\mr{\theta}^{(\ell)}}

\newcommand{\loplus}{%
\begin{tikzpicture}%
\hspace{1.5pt} \draw (0,0) circle (3.5pt);
\draw (0,3.5pt) -- (0,-3.5pt);
\draw (0,0) -- (3.5pt,0); 
\end{tikzpicture}
\hspace{3pt}
}

\title{Near-Horizon Symmetries in Einstein-Maxwell theory}
\author[a,b]{Gianfranco De Simone \orcid{0009-0004-5342-7670} ,}
\affiliation[a]{Università degli Studi di Udine,\\ via Palladio 8, I-33100 Udine, Italy}
\affiliation[b]{National Institute for Nuclear Physics (INFN), Sezione di Trieste,\\ Via Valerio 2, 34127, Italy}
\emailAdd{gianfranco.desimone@uniud.it}

\abstract{This manuscript aims to provide a comprehensive derivation of the Einstein-Maxwell charges and fluxes in the near-horizon region of a four-dimensional non-extremal black hole, with vanishing cosmological constant. Specifically, we present a detailed derivation of the Noether charges within both the metric and first-order formulations, elucidating the relationship between the Carrollian internal boost charge and the Lorentz boost charge. It is well-established in the literature that Carrollian fluids exhibit an internal local boost symmetry; we demonstrate that this symmetry precisely corresponds to a Lorentz internal transformation. Finally, we prove that the near-horizon Einstein equations can be obtained from the flux-balance law by employing the generalized Barnich-Troessaert bracket.}

\begin{document}

\maketitle
\flushbottom

\section{Introduction}
In the early sixties, Bondi, Metzner
and van der Burg in \cite{Bondi:1960jsa,Bondi:1962px}, and thereafter Sachs in \cite{Sachs:1961zz, Sachs:1962wk}, by analysing the geometry of an asymptotically flat spacetime (AFS) in presence of gravitational radiation, showed that the gravitational symmetry group at the future null infinity $\cI^+$ consists of an infinite dimensional group, dubbed as the BMS group (see \cite{Madler:2016xju} for a review). In the last decade, renewed interest in the BMS group has emerged due to the discovery of a surprising  relationship between asymptotic symmetries \cite{Bondi:1960jsa,Bondi:1962px, Sachs:1961zz, Sachs:1962wk, Newman:1962cia}, Weinberg soft theorems \cite{Weinberg:1965nx} and memory effects \cite{Christodoulou:1991cr,Braginsky:1987kwo}, whose connections are encapsulated  in the so-called infrared triangle \cite{Strominger:2013jfa, Strominger:2014pwa} (see \cite{Strominger:2017zoo} for a review). These remarkable results paved the way for a wealth of new developments, including the celestial holography proposal which aims to describe a theory in an asymptotically flat spacetime as a conformal field theory defined on the celestial sphere $\cC\cS$. This correspondence establishes a map between four-dimensional scattering amplitudes and correlation functions in a two-dimensional conformal field theory \cite{Strominger:2017zoo, Donnay:2020guq, Kapec:2014opa, Kapec:2016jld, Pasterski:2016qvg}.
Alongside the celestial holography project, a new dual description for 4-dimensional AFS has been proposed, known as Carrollian holography, where the dual theory consists of a conformal (Carrollian) field theory lying on the asymptotic 3-dimensional null hypersurface. A relationship between these two descriptions have been explored in \cite{Ciambelli:2018wre,Ciambelli:2018ojf,Setare:2018ziu,Ciambelli:2019lap,Donnay:2022aba,Donnay:2022wvx,Bagchi:2022emh}.\\
Simultaneously, BMS-like symmetries have been explored for null hypersurfaces at finite distance \cite{Chandrasekaran:2018aop, Hopfmuller:2018fni, Adami:2021nnf,Adami:2021kvx}, with a particular focus on black hole's horizon \cite{Donnay:2015abr, Donnay:2016ejv, Adami:2020amw, Liu:2022uox, Sheikh-Jabbari:2022mqi, Chandrasekaran:2023vzb}
(and also non-expanding horizons \cite{Ashtekar:2021kqj, Ashtekar:2021wld, Ashtekar:2025wnu}), leading to the soft-hair proposal in the context of black hole's information paradox \cite{ Hawking:2016msc, Hawking:2016sgy, Haco:2018ske, Mao:2016pwq,Grumiller:2019fmp}. A significant insight into understanding the near-horizon geometry was provided in \cite{Donnay:2019jiz} (see also \cite{Penna:2018gfx,Freidel:2022vjq, Adami:2023wbe, Freidel:2024emv}), where the authors demonstrated the emergence of a Carrollian physics at the horizon when the ultra-relativistic limit for the stretched horizon \cite{MacDonald:1982zz, Thorne:1987bsa, Price:1986yy} is taken. \\
A connection between finite-distance null hypersurfaces and asymptotic null boundaries was investigated in \cite{Ciambelli:2025mex}, where the author demonstrated that the finite-distance gravitational phase space asymptotes to the Ashtekar-Streubel \cite{Ashtekar:1981bq} phase space. Moreover, he also showed that the Bondi mass loss formula can be recovered by evaluating the sub-sub-leading order of the null Raychaudhuri equation as the position of the null hypersurface approaches to infinity.\\
In this work we continue the investigation of the near horizon symmetries, providing a comprehensive derivation of the charges and fluxes near a four-dimensional non-extremal black hole in presence of electro-magnetic radiation. Our results suggest that a non-trivial electric Noether charge appears at the sub-leading order, while at leading order it gives a vanishing contribution. Moreover, we furnished a charge analysis in both metric and tetrad formulation, highlighting a connection between the Carrollian internal boost charge and the Lorentz boost charge. Indeed, as noticed in \cite{Ciambelli:2023mir}, Carrollian fluids possess an internal local boost symmetry, whose charge is non-vanishing and equals to the corner area element. In appendix \ref{EC}, we compute the value of the Lorentz charge in the Einstein-Cartan formulation of gravity, and observe that acts exactly as the Carrollian internal local boost. As argued in \cite{Ciambelli:2023mir}, this local boost charge provides a notion of
entropy and could be used as a candidate to describe the generalized entropy (see also the recent work \cite{Shajiee:2025cxl} where the authors emphasize that gravitational entropy is regarded as the charge associated with the local boosts).

\noindent
The paper is organized as follows: the first section provides an overview of the covariant phase space formalism used to compute the charge content of the theory \cite{Wald:1999wa, Harlow:2019yfa, Speranza:2017gxd, Chandrasekaran:2020wwn, Donnelly:2016auv, Freidel:2021cjp, Barnich:2001jy}. In section \ref{sec2} we discuss the geometrical properties of null hypersurfaces embedded into a 4-dimensional spacetime manifold \cite{Gourgoulhon:2005ng, Booth:2005ng, Booth:2006bn, Booth:2012xm, Ciambelli:2023mir, Chandrasekaran:2021hxc}  and solve the Einstein equations for the near-horizon metric ansatz \eqref{metric}. By integrating the Einstein and the Maxwell hypersurface equations, i.e., $\mathbb{E}_{\rho\mu}=0$ and $\mathbb{M}_\rho=0$ respectively, we obtain the near-horizon expansions of the metric and the Maxwell potential. Then, we explicitly write down up to the second order in radial coordinate $\rho$ the behaviour of the metric and the dual Maxwell tensor. In section \ref{sec3} we compute the canonical and the Einstein-Hilbert Noether charges and fluxes, and subsequently analyse the charge algebra via the Barnich-Troessaert bracket \cite{Barnich:2009se,Barnich:2010eb,Barnich:2011mi,Barnich:2013axa}. In particular, we demonstrate that the null Raychaudhuri and Damour equations are derived holographically via the flux-balance law related to the generalized Barnich-Troessaert bracket \cite{Freidel:2021cjp}.
Moreover, through the computation performed in appendix \ref{EC}, we show that the internal local boost symmetry identified in \cite{Ciambelli:2023mir} corresponds to an internal Lorentz symmetry.\\
The charge calculation within the Einstein-Cartan formulation and the flux-balance laws can be seen as consistency checks for our analysis.

\vspace{0.3cm}

\noindent
\emph{Notation and conventions}:
We use the metric signature ($-+++$) and work in natural units $8\pi G=c=e=1$. We denote the spacetime indices by Greek letters, $\mu, \nu, \sigma,\cdots$; the indices on the null horizon by lowercase Latin letters from the middle of the alphabet $i,j,k,\cdots$; the indices on the spacelike codimension-2 surface $\cS$ by lowercase Latin letters from the beginning of the alphabet  $a,b,c,\cdots$. The notation for (anti-) symmetrizing indices is 
\begin{equation*}
O_{(\mu}P_{\nu)}= \f{1}{2}(O_\mu P_\nu + O_\nu P_\mu)\qquad (\text{resp.}\  O_{[\mu}P_{\nu]}= \f{1}{2}(O_\mu P_\nu-O_\nu P_\mu))     
\end{equation*}
and by $O_{\langle \mu\nu \rangle}$ we denote the symmetric trace-free part of $O_{\mu\nu}$. The volume form reads as follows
\begin{equation*}
    \eps = \f{1}{4!} \sqrt{-g}\  \epsilon_{\mu\nu\rho\sigma}\ \dext x^\mu\wedge \dext x^\nu \wedge \dext x^\rho\wedge \dext x^\sigma
\end{equation*}
with $\epsilon_{v\rho \theta\phi}=1$. The on-shell symbol is denoted by $\heq$ and the equalities that hold at the horizon (that is, when $\rho=0$) are denoted by $\nueq$. Moreover, the values assumed by certain quantities on the horizon are customized by a circle on the top, \ie $\mathring{a}\nueq a$. 
The exterior derivative and the interior product with respect to a generic vector field $\xi$ are denoted by $\dext$ and $\iota_\xi$, respectively.

\section{A quick review of CPS}\label{CPS}
In this section, we briefly review the charge prescription used in this work, referring the reader to the comprehensive treatments presented in the following references \cite{Wald:1999wa,Donnelly:2016auv,Freidel:2020xyx,Freidel:2021cjp,Speranza:2017gxd,Harlow:2019yfa,Chandrasekaran:2020wwn,Barnich:2001jy} for further details. Given a $n$-form Lagrangian $\bold{L}$, one can uniquely define a $(n-1)$-form $\bd{\theta}$, known as the pre-symplectic potential, such that 
\begin{equation}
    \var \bold{L} \heq \dext \bd{\theta},
\label{varL=dtheta}
\end{equation}
where the symbol $\heq$ indicates equality on-shell. The pre-symplectic potential encodes the symmetry structure of the theory and plays a central role in the construction of conserved charges. Given a vector field $\xi$, the associated Noether charge aspect $\bd{q}_\xi$ is defined through the following relation
\begin{equation}
\dext\bd{q}_\xi := I_\xi\bd{\theta} - \iota_\xi \bold{L}-\bd{a}_\xi, 
\label{dq_xi}
\end{equation} 
where $\bd{a}_\xi = \Delta_\xi \bold{L}$  represents the anomaly of the Lagrangian under the symmetry transformation generated by $\xi$. By definition, for an arbitrary operator $\cO$ and a vector field $\xi$, the anomaly $\Delta_\xi\cO$ is given by
\begin{equation}
\Delta_\xi\cO := \var_\xi \cO - \Lie_\xi\cO -I_{\var\xi} \cO, 
\label{anom_def}
\end{equation}
where $I_{\var\xi}$ accounts for the field-dependence of the vector field $\xi$ \cite{Donnelly:2016auv, Speranza:2017gxd, Freidel:2021cjp}. Then, the Noether charge $\cQ_\xi$ is obtained by integrating the equation \eqref{dq_xi} over a codimension-1 hypersurface $\Sigma$, and it reads
\begin{equation}
    \cQ_\xi := \int_{\pa\Sigma=\cS} \bd{q}_\xi \label{charge_exp},
\end{equation}
where $\cS$ is a codimension-2 surface, known as the corner. In addition to the Noether charge, it is necessary to define the so-called Noetherian flux \cite{Freidel:2021cjp}, which accounts for the potential leakage of symplectic information through the corner $\cS$. This Noetherian flux is given by
\begin{equation}
\cF_\xi :=\int_\cS (\iota_\xi\bd{\theta} +\bd{A}_\xi) + \int_\cS \bd{q}_{\delta\xi}\label{flux},
\end{equation}
where $\bd{A}_\xi$ is the \emph{symplectic anomaly}, whose exterior derivative is expressed as
\begin{equation}
\dext \bd{A}_\xi :=\Delta_\xi \bd{\theta}-\delta \bd{a}_\xi +\bd{a}_{\var\xi}.
\label{symanom}
\end{equation}
The charge algebra is then defined via the generalized Barnich-Troessaert bracket \cite{Freidel:2021cjp, Barnich:2011mi}
\begin{equation}
    \{\cQ_\xi, \cQ_\zeta\}_L := \var_\xi \cQ_\zeta - I_\zeta \cF_\xi + \cK_{(\xi, \zeta)},\label{Q_algebra}
\end{equation}
where $\cK_{(\xi, \zeta)}$ is a 2-cocycle given by
\begin{equation}
\cK_{(\xi, \zeta)}:=\int_{\cS}\iota_\xi\iota_\zeta \bold{L} + \int_\cS(\iota_\xi \bd{a}_\zeta-\iota_\zeta \bd{a}_\xi)+\int_\cS \bd{c}_{(\xi,\zeta)} \label{cocy}.
\end{equation}
Here, $\bd{c}$ is a codimension 2-form, whose exterior derivative is (assuming the lemma in sec. 3.2 of \cite{Freidel:2021cjp} holds)
\begin{equation}
\dext \bd{c}_{(\xi,\zeta)} := \Delta_\xi\bd{a}_\zeta-\Delta_\zeta\bd{a}_\xi +\bd{a}_{[\![ \xi, \zeta ]\!]}\label{cexpr},
\end{equation}
where the double bracket notation represents the modified Lie bracket \cite{Barnich:2010eb}, defined as
\begin{equation}  
[\![ \xi, \zeta ]\!] := [\xi, \zeta]_{\text{Lie}} +\delta_\zeta \xi -\delta_\xi \zeta.
\label{mod_lie}
\end{equation}
At the beginning, we claimed that the pre-symplectic potential is uniquely determined from the Lagrangian. Hence, under a Lagrangian shift $\bold{L} \to \bold{L}'=\bold{L}+\dext \bd{\ell}_{\cB}$, the pre-symplectic potential of the shifted Lagrangian reads as
\begin{equation}
    \bd{\theta} \to \bd{\theta}'=  \bd{\theta} + \var\bd{\ell}_\cB - \dext\bd{\vartheta},
\end{equation}
where $\bd{\vartheta}$ is the corner potential. The
Noether charge and flux then change as follows
\begin{equation}
\cQ_\xi' = \cQ_\xi + \int_S (\iota_\xi\bd{\ell}_\cB -I_\xi \bd{\vartheta}).\qquad
\cF_\xi' = \cF_\xi +\int_S (\delta\iota_\xi\bd{\ell}_\cB-\delta_\xi\bd{\vartheta})\label{mod_Q}
\end{equation}
and the shifted 2-cocycle is
\begin{equation}
\cK'_{(\xi, \zeta)}= \cK_{\xi,\zeta} + \int_S \iota_\xi\iota_\zeta \dext\bd{\ell}_\cB 
+ \int_S (\iota_\xi\Delta_\zeta \bd{\ell}_\cB-\iota_\zeta \Delta_\xi\bd{\ell}_\cB).\label{mod_K}
\end{equation}
Notably, the off-shell flux balance law
\begin{equation}
     \{\cQ_\xi, \cQ_\zeta\}_L - \var_\xi\cQ_\zeta + I_\zeta \cF_\xi - \cK_{(\xi, \zeta)} = -\int_\cS\iota_\xi \bd{C}_\zeta\label{fl_law_BT},
\end{equation}
is invariant under Lagrangian shift as demonstrated in \cite{Freidel:2021cjp}, i.e.,
\begin{equation}
     \{\cQ_\xi, \cQ_\zeta\}_L + \cQ_{[\![\xi, \zeta]\!]} = \{\cQ_\xi, \cQ_\zeta\}_{L'} + \cQ'_{[\![\xi, \zeta]\!]}.
\end{equation}

\section{Near-horizon geometry}\label{sec2}
In this section, we outline the construction of the spacetime geometry in the neighbourhood of a null horizon \cite{Booth:2005ng, Booth:2006bn, Booth:2012xm, Krishnan:2012bt}. The first part briefly reviews some concepts related to the geometry of null hypersurfaces, referring the reader to the following references \cite{Ashtekar:2025wnu, Chandrasekaran:2021hxc, Gourgoulhon:2005ng} for a detailed treatment. In the final part, using the metric ansatz \eqref{metric}, we solve the Einstein and Maxwell equations, and present the metric expansion up to the second order in the radial coordinate.

\subsection{Null geometries}\label{nullhyper}
In this subsection, we recall some basic definitions and properties of null hypersurfaces. Let us begin by considering a null hypersurface $(\cN, \bd{h})$ embedded into a spacetime manifold $(\cM, \bd{g})$ via the embedding map 
\begin{equation}
    \Pi : \cN \to \cM.
\end{equation}
Assume that the topology of our null hypersurface is $\cN \simeq \cS \times \R$, where $\cS$ is a compact codimension-2 hypersurface endowed with a spatial metric $q_{ab}$. Let $\bd{\ell}$ denote a future-directed null normal to $\cN$, defined up to a local rescaling
\begin{equation}
    \ell_\mu \to e^\lambda \ell_\mu,\label{ldef}
\end{equation}
where $\lambda(v,x)$ is a smooth function on $\cN$. Given the spacetime metric and the null normal to $\cN$, we can define the null vector $\ell^\mu = g^{\mu\nu}\ell_\nu$, which defines the integral curves generating $\cN$. Thus, we can write $T\cN = \{\ell^i\}$. In particular, $\ell^\mu$ is the push-forward of $\ell^i$ in $T\cM$. Once $\ell$ is specified, the non-affinity parameter $\kappa$ is defined by
\begin{equation}
    \ell^\mu\nabla_\mu \ell^\nu = \kappa \ell^\nu,
\end{equation}
and we write $\mr{\kappa}\nueq \kappa$ to denote its value on $\cN$. Subsequently, we define a pull-back map $\Pi^*$ which sends a $p$-form in $T^*\cM$ to a $p$-form in $T^*\cN$, i.e.,
\begin{equation}
    \omega_\mu \to \Pi^\mu_{\ i} \omega_\mu \qquad \text{for}\  \bd{\omega}\in T^*\cM,
\end{equation}
and $\Pi^\mu_{\ i}\ell_\mu=0$. In particular, given a contraction in $\cM$, say $v^\mu w_\mu$, the vector $v^i$ is well defined on $\cN$ if and only if $v^\mu\ell_\mu=0$, so that $v^\mu w_\mu= v^i w_i$. Conversely, if we have a contraction $v^i w_i$ on $\cN$, the 1-form $w_a$ is well-defined on $\cS$ if and only if $w_i \ell^i=0$, so that $v^i w_i= v^a w_a$.
Using the embedding map, the induced metric on $\cN$ is given by
\begin{equation}
    h_{ij}=\Pi^\mu_{\ i}\Pi^\nu_{\ j} g_{\mu\nu},
\end{equation}
and is degenerate, $h_{ij}\ell^j=0$. From the above discussion, we can regard the metric $h_{ij}$ as a 2-tensor defined on $T^*\cS \otimes T^*\cS$, i.e., as $h_{ab}=q_{ab}$. Next, we define the second fundamental form of $\cN$ as
\begin{equation}
    K^{(\ell)}_{ij}= \Pi^\mu_{\ i}\Pi^\nu_{\ j} \nabla_\mu\ell_\nu,
\end{equation}
and since $\ell^i K^{(\ell)}_{ij}=0$, we can write $K^{(\ell)}_{ij}=K^{(\ell)}_{ab} \in T^*\cS \otimes T^*\cS$. It is also instructive to observe that the pullback on $\cN$ of the covariant derivative of $\ell$, 
\begin{equation}
    \Pi^\mu_{\ i}\nabla_\mu \ell^\nu,
\end{equation}
is an intrinsic tensor of $\cN$. This follows because $\ell_\nu\Pi^\mu_{\ i}\nabla_\mu \ell^\nu = \Pi^\mu_{\ i}\nabla_\mu (\ell_\nu\ell^\nu)/2=0$. This quantity is called \emph{shape operator} or \emph{Weingarten map} \cite{Chandrasekaran:2018aop,Gourgoulhon:2005ng},
\begin{equation}
    \cW_i^{\ j} := \Pi^\mu_{\ i}\nabla_\mu \ell^\nu\Pi^{j}_{\ \nu},\label{Wdef}
\end{equation} 
and describes how $\cN$ bends in $\cM$. It acts as an endomorphism of $\cN$, associating to each vector $v\in T\cN$ the vector $v\to \nabla_v \ell$. The explicit expression of the projector $\Pi$ is determined once an auxiliary 1-form $\bd{n}$, satisfying $\ell\cdot n=-1$, is introduced. The projector then reads
\begin{equation}
    \Pi^\mu_{\ i} = \var_{\ i}^{\mu} +n^\mu \ell_i,\label{20}
\end{equation}
with $\Pi^\mu_{\ i}\Pi^i_{\ \nu}=\Pi^\mu_{\ \nu}$. The spacetime metric then is
\begin{equation}
    g_{\mu\nu}= q_{\mu\nu} - \ell_\mu n_\nu - n_\mu\ell_\nu.
\end{equation}
Moreover, an induced (rigged) connection on $\cN$ can be defined via the projector $\Pi$ \cite{Mars:1993mj,Gourgoulhon:2005ng}. Given a vector $v\in T\cN$, the rigged covariant derivative is defined as
\begin{equation}
    D_i v^j = \Pi_{\ i}^\mu \Pi^j_{\ \nu} \nabla_\mu v^\nu,
\end{equation}
and depends on the choice of $\bd{n}$. From \eqref{20} and the definition of the rigged covariant derivative, one obtains the following relation 
\begin{equation}
D_i \eps_\cN = -\omega_i \eps_\cN,
\label{D_rel}
\end{equation}
where $\omega_i$ is the rotation 1-form and its spatial projection
\begin{equation}
    \pi_a = - q^\mu_{\ a} n_\nu\nabla_\mu \ell^\nu \label{haji}
\end{equation}
is the Hajicek field. As shown in \cite{Chandrasekaran:2021hxc}, a Brown-York tensor associated with a null hypersurface can be defined via the Weingarten operator in \eqref{Wdef} as follows
\begin{equation}
    T_i^{\ j} = \cW_i^{\ j} - \cW \var_i^{\ j}.
\label{Tij}
\end{equation}
Using the explicit form of the projector \eqref{20}, the Weingarten operator is
\begin{equation}
    \cW_i^{\ j} = K^{(\ell)j}_{i} +\ell^j \omega_i
\end{equation}
and its trace is $\cW =\theta^{(\ell)} +\kappa$. 
Finally, we argued that the null normal to $\cN$ is not uniquely defined because of the scaling transformation in \eqref{ldef}. Consequently, the various quantities defined so far transform under the rescaling \eqref{ldef} as follows \cite{Chandrasekaran:2018aop}
\begin{equation}
\begin{aligned}
K^{(\ell)}_{ij}\to e^{\lambda}K^{(\ell)}_{ij},\qquad
\cW_i^{\ j} \to e^\lambda (\cW_i^{\ j} +D_i\lambda\ \ell^j),\qquad
\kappa \to e^\lambda (\kappa +\Lie_\ell \lambda).
\end{aligned}
\end{equation}

\subsection{Near-horizon metric}
In the preceding subsection, we outlined some important properties of null hypersurfaces that will be useful in the subsequent analysis. In this subsection, we introduce the near-horizon geometry, referring the reader to treatment presented in \cite{Booth:2006bn, Booth:2012xm} for further details (see \cite{Krishnan:2012bt} for a description of isolated horizon within the Newman-Penrose formalism). Let $\cN$ be defined as a smooth union of $(n-2)$ spacelike hypersurfaces $\cS_v$, i.e., $\cN = \bigcup_v \cS_v$, where $v$ is the advanced time coordinate. Then, we choose a coordinate system $\sigma^a=(\theta,\phi)$ on $\cS$, and extend these coordinates onto $\cN$ through the vector field $\ell$. The normal space to $\cS$ is spanned by the future-oriented null vectors $\ell^\mu$ and $n^\mu$, satisfying $\ell\cdot n=-1$. These vectors are defined up to the scaling symmetry \eqref{ldef}
\begin{equation}
    \ell \to e^\lambda \ell, \qquad n\to e^{-\lambda} n.
\end{equation}
Finally, the coordinate system is extended off $\cN$ using the vector field $n^\mu = \pa_\rho$. The near-horizon metric can be expressed as follows \cite{Booth:2012xm}
\begin{equation}
    \dext s^2 = -2\dext v\dext \rho + 2 V\dext v^2 +q_{ab}(\dext \sigma^a+ U^a\dext v)(\dext \sigma^b + U^b \dext v),
    \label{metric}
\end{equation}
where $V, U^a$ and $q_{ab}$ depend on the coordinates $(v,\rho, \sigma^a)$, and the inverse metric is given by
\begin{equation}
    g^{\mu\nu}\pa_\mu\pa_\nu = -2\pa_v\pa_\rho -2V\pa_\rho\pa_\rho +2U^a\pa_a\pa_\rho +q^{ab}\pa_a\pa_b.
\end{equation}
The near-horizon metric \eqref{metric} is written in the so called Newman-Unti gauge, whose gauge conditions are
\begin{equation}
    g_{v\rho}=-1, \qquad g_{\rho \rho}=0, \qquad g_{\rho a}=0,\label{NU}
\end{equation}
and we impose the following behaviour of the metric at the boundary 
\begin{equation}
    g_{vv}=O(\rho), \qquad g_{va}=O(\rho), \qquad g_{ab}=O(1).
    \label{BC}
\end{equation}
From the metric \eqref{metric}, we have that the two null vectors normal to $\cN$ are
\begin{equation}
    \ell^\mu \pa_\mu =  \pa_v +V\pa_\rho- U^a \pa_a, \qquad n^\mu= \pa_\rho
\end{equation}
and the relative 1-forms read
\begin{equation}
    \bd{\ell}=-\dext \rho  + V \dext v, \qquad \bd{n}= -\dext v.
\end{equation}
At this point, we can compute the extrinsic curvatures
\begin{equation}
    K^{(\ell)}_{ab}= \f{1}{2}q^{\mu}_{\ a}q^{\nu}_{\ b}\Lie_\ell q_{\mu\nu},\qquad \text{and} \qquad     K^{(n)}_{ab}= \f{1}{2}q^{\mu}_{\ a}q^{\nu}_{\ b}\Lie_n q_{\mu\nu},
\end{equation}
which explicitly read
\begin{equation}
    K^{(\ell)}_{ab} = \f{1}{2}\Bigl(\pa_v q_{ab} +V\pa_\rho q_{ab} - 2D_{(a}U_{b)}\Bigl), \qquad\text{and}\qquad
    K^{(n)}_{ab} = \f{1}{2} \pa_\rho q_{ab}.
\end{equation}
Using the definition in \eqref{haji}, the components of the rotational 1-form are
\begin{equation}
    \omega_a = \f{1}{2}q_{ab}\pa_\rho U^b, \qquad \omega_v = \pa_\rho V + \f{1}{2}U^a q_{ab}\pa_\rho U^b,
\end{equation}
and the in-affinity parameter is
\begin{equation}
    \kappa = \pa_\rho V,
\end{equation}
where one can easily check the relation $\omega_\mu \ell^\mu = \kappa$. In particular, the metric functions in \eqref{metric} are determined via the in-affinity parameter and the Hajicek field by means of the following relations
\begin{equation}
V = \int \dext\rho\ \kappa, \qquad \text{and}\qquad U^a =2 \int \dext\rho\ \pi^a .\label{FU}
\end{equation}

\subsection{Einstein-Maxwell theory}
Having at hand the ansatz \eqref{metric} describing the metric near-horizon, we can plug the latter into the Einstein equations and solve them order by order in the radial coordinates. The Einstein-Maxwell (EM) Lagrangian form in a four-dimensional spacetime reads
\begin{equation}
    \bold{L}_\EM[\bd{g}, \bd{A}] = \Bigl(\f{1}{2}R-\f{1}{4} F_{\mu\nu}F^{\mu\nu}\Bigl)\eps,
\end{equation}
where $\eps= \sqrt{-g}\ \dext^4 x$ is the volume form and $F_{\mu\nu} = \pa_\mu A_\nu - \pa_\nu A_\mu$ is the electromagnetic field strength. By varying the EM Lagrangian with respect to the metric field $g_{\mu\nu}$ and the gauge field $A^\mu$, we obtain the Einstein equations
\begin{equation}
    \mathbb{E}_{\mu\nu} := R_{\mu\nu}-\f{1}{2}Rg_{\mu\nu} - 2 T_{\mu\nu}=0,
    \label{Eins_eq}
\end{equation}
where
\begin{equation}
    T_{\mu\nu} = F_{\mu\sigma}F^{\sigma}_{\ \nu} -\f{1}{4}g_{\mu\nu}F_{\alpha\beta}F^{\alpha\beta}
\end{equation}
is the Maxwell stress-energy tensor, and the Maxwell field equations
\begin{equation}
    \mathbb{M}_{\nu} := \nabla^\mu F_{\mu\nu}=0,
\end{equation}
respectively. In 4-dimensions, the Maxwell stress-energy tensor is traceless and therefore by tracing the \eqref{Eins_eq} we obtain $R=0$. 
Let us work in the radial gauge $A_\rho=0$ and assume the following radial expansions for the spacelike metric field
\begin{equation}
    q_{ab}(v,\rho, \sigma^c) = \mr{q}_{ab}(v,\sigma^c) + \rho \lambda_{ab}(v,\sigma^c) +\rho^2 d_{ab}(v,\sigma^c) + O(\rho^3),
\end{equation}
and for angular components of Maxwell potential
\begin{equation}
    A_a(v,\rho, \sigma^c) = \mr{A}_a(v,\sigma^c) +\rho B_a(v,\sigma^c) + \rho^2 B^{(1)}_a(v,\sigma^c)+ O(\rho^3),
\end{equation}
where $\mr{q}_{ab}, \lambda_{ab}, d_{ab},\mr{A}_a, B_a,$ and $ B^{(1)}_a$ depend on $(v,\sigma^c)$. The condition $q^{ac}q_{bc}=\var^a_{\ b}$ implies
\begin{equation}
\begin{aligned}
    q^{ab} &= \mr{q}^{ab} -\rho \lambda^{ab} -\rho^2 (\sc{d}^{ab} - \lambda^{bc}\lambda^a_{\ c})+ o(\rho^3),
\end{aligned}
\end{equation}
and let us split the horizontal covariant derivative $D_a = q^i_{\ a} D_i$ into leading and sub-leading contributions, as
\begin{equation}
    D_a U^b = \mr{D}_a U^b + \fr{C}^b_{ac}U^c,
\end{equation}
where
\begin{equation}
    \fr{C}^b_{ac}=\f{\rho}{2}\mr{q}^{bd}(\mr{D}_a \lambda_{dc} +\mr{D}_c \lambda_{da} -\mr{D}_d \lambda_{ac})+O(\rho^2).
    \label{chrC}
\end{equation}
In order to find the near-horizon behaviour of the metric field, we have to solve the hypersurface equations
\begin{equation}
\begin{aligned}
    \mathbb{E}_{\rho \mu} &:= R_{\rho \mu} - \f{1}{2}Rg_{\rho\mu} - 2 T_{\rho\mu}=0\\
\end{aligned}
\end{equation}
to determine the radial expansion of the metric field, and
\begin{equation}
    \mathbb{M}_\rho:= \nabla^\mu F_{\mu\rho}=0
\end{equation}
for accessing to the radial behaviour of the Maxwell field.

\subsubsection{Radial Einstein equations}
In this subsection we derive and solve the hypersurface Einstein's equations and write down the metric radial-expansion up to the second order in $\rho$. The radial components of the Ricci tensor are
\begin{equation}
\begin{aligned}
R_{\rho\rho} &=  K^{(n)b}_{a}K^{(n)a}_{b}-  q^{ab}\pa_\rho K^{(n)}_{ab},\\ 
R_{a\rho}  &= D_b K^{(n)b}_{a}- D_a \theta^{(n)} - \theta^{(n)} \pi_a -\pa_\rho\pi_a, \\
R_{v\rho} &= -\kappa \theta^{(n)}- \pa_\rho \kappa + \f{1}{2} K^{ab}_{(n)} \pa_v q_{ab} - q^{ab} \pa_v K^{(n)}_{ab}  +U^a D_a\theta^{(n)}- 2\pi_a \pi^a + D_a \pi^a\\
& \qquad + K^{(n)b}_a D_b U^a ,
\end{aligned}
\label{EEs}
\end{equation}
and the radial components of the stress-energy tensor read
\begin{equation}
\begin{aligned}
    T_{\rho\rho} &= -q^{ab}\pa_\rho A_a \pa_\rho A_b,\\
    T_{\rho a} &= q^{cb}F_{ba}\pa_\rho A_c + U^{b}\pa_\rho A_b \pa_\rho A_a  - \pa_\rho A_v\pa_\rho A_a ,\\
    T_{\rho v} &= q^{ab} \pa_\rho A_a (\pa_b A_v-\pa_v A_b) + U^a\pa_\rho A_a \pa_\rho A_v - (\pa_\rho A_v)^2 +\f{1}{4}|F|^2.
\end{aligned}
\end{equation}
The first hypersurface equation constraints the trace of the boundary metric, i.e.
\begin{equation}
    \mathbb{E}_{\rho\rho} =  K^{(n)b}_{a}K^{(n)a}_{b}-  q^{ab}\pa_\rho K_{ab}^{(n)} +2q^{ab} \pa_\rho A_b \pa_\rho A_a,
\end{equation}
and yields the following radial behaviour for the transversal expansion
\begin{equation}
    \theta^{(n)}(v,\rho, \sigma^c) =\mr{\theta}^{(n)}(v,\sigma^c) + \rho( \mr{K}_a^{(n)b}\mr{K}^{(n)a}_b +2 B_a B^a) + o(\rho).
\end{equation}
The $\rho a$-component of the Einstein tensor is
\begin{equation}
\begin{aligned}
    \mathbb{E}_{\rho a} &= D_b K^{(n)b}_{a}- D_a \theta^{(n)} - \theta^{(n)} \pi_a -\pa_\rho\pi_a  -2q^{bc} F_{ba}\pa_\rho A_c\\
    &\quad + 2\pa_\rho A_v \pa_\rho A_a -2U^b \pa_\rho A_b \pa_\rho A_a,
\end{aligned}
\end{equation}
and determine the radial expansion of the Hajicek field, 
\begin{equation}
    \pi_a(v,\rho, \sigma^c) = \mr{\pi}_a (v,\sigma^c) + \rho\Bigl( \mr{D}_b\mr{K}_a^{(n)b} - (\mr{D}_a +\mr{\pi}_a)\mr{\theta}^{(n)} - 2\mr{F}^{\ b}_{a}B_b \Bigl) +o(\rho).
\end{equation}
Lastly, the $\rho v$-component of the Einstein tensor reads as
\begin{equation}
\begin{aligned}
    \mathbb{E}_{\rho v}  &= -\kappa \theta^{(n)}- \pa_\rho \kappa + \f{1}{2} K^{ab}_{(n)} \pa_v q_{ab} - q^{ab} \pa_v K^{(n)}_{ab}  +U^a D_a\theta^{(n)}- 2\pi_a \pi^a+ D_a \pi^a\\
    &  + K^{(n)b}_a D_b U^a + \f{1}{2}(R -\abs{F}^2)-2q^{ab}\pa_\rho A_a F_{bv} +2(\pa_\rho A_v)^2  -2 U^a \pa_\rho A_a \pa_\rho A_v,
\end{aligned}
\end{equation}
and gives the radial expansion of the in-affinity parameter,
\begin{equation}
\begin{aligned}
    \kappa(v,\rho, \sigma^c) &= \mr{\kappa}(v, \sigma^c) + \rho\Bigl(-\mr{K}^{ab}_{(n)}\mr{K}^{(\ell)}_{ab} - (\pa_v + \mr{\kappa})\mr{\theta}^{(n)} + (\mr{D}_a -2\mr{\pi}_a)\mr{\pi}^a\\
    &+2 B^a\pa_v\mr{A}_a + \f{1}{2}(\mr{R}- |\mr{F}|^2)  \Bigl) + o(\rho).
\end{aligned}
\end{equation}
Now, let us compute the $\mathbb{M}_\rho$ component of the Maxwell equation, which gives information about the radial expansion of $A_v$ in terms of the angular components $A_a$. We obtain
\begin{equation}
\begin{aligned}
 \mathbb{M}_\rho &= (\pa_{\rho} + \theta^{(n)}) \pa_\rho A_v - U^{a}(\pa_{\rho} +\theta^{(n)} )\pa_\rho A_a -(D^a + 2\pi^a)\pa_\rho A_a ,
\end{aligned}
\end{equation}
and yields
\begin{equation}
    A_v= \f{\rho^2}{2}(\mr{D}_a+2\mr{\pi}_a) B^a  +o(\rho^2).
\end{equation}

\noindent
Imposing the boundary conditions in \eqref{BC} and using the above expansions, the metric functions in \eqref{FU} read
\begin{equation}
\begin{aligned}
V&= \rho\mr{\kappa} + \f{\rho^2}{2} \Bigl(-\mr{K}^{ab}_{(n)}\mr{K}^{(\ell)}_{ab} - (\pa_v + \mr{\kappa})\mr{\theta}^{(n)} + (\mr{D}_a -2\mr{\pi}_a)\mr{\pi}^a + \f{1}{2}(\mr{R}- |\mr{F}|^2)\\
&\quad +2 B^a\pa_v\mr{A}_a \Bigl) + o(\rho^2),\\
U^a&= 2\rho\mr{\pi}^a + \rho^2\Bigl(\mr{D}_b\mr{K}^{ab}_{(n)} - (\mr{D}^a + \mr{\pi}^a) \mr{\theta}^{(n)} -2 \mr{F}^{ab} B_b- 2\mr{K}_{(n)}^{ab}\mr{\pi}_b\Bigl) + o(\rho^2),\\ 
q_{ab} &= \mr{q}_{ab} + 2\rho\mr{K}^{(n)}_{ab}+ \rho^2 \Bigl(d_{\langle ab \rangle} +\f{1}{2} \mr{q}_{ab}(|\mr{K}^{(n)}|^2 +2 |B|^2)
\Bigl) + o(\rho^2).
\end{aligned}
\end{equation}
Hence, the metric up to the second order in the radial coordinate reads 
\begin{equation}
\begin{aligned}
    \dext s^2 &= -2\dext v\dext \rho + \mr{q}_{ab} \dext \sigma^a \dext \sigma^b \\
    &\quad +2\rho\Bigl\{ \mr{\kappa}\dext v^2 
    +2\mr{\pi}_a \dext \sigma^a \dext v
    + \mr{K}^{(n)}_{ab}\dext \sigma^a \dext \sigma^b\Bigl\}\\
    & \quad + \rho^2 \Bigl\{\Bigl(-\mr{K}^{ab}_{(n)}\mr{K}^{(\ell)}_{ab} - (\pa_v + \mr{\kappa})\mr{\theta}^{(n)} + (\mr{D}_a +2\mr{\pi}_a)\mr{\pi}^a+ \f{1}{2}(\mr{R}-|\mr{F}|^2) \\
    &\quad +2 B^a\pa_v\mr{A}_a \Bigl)
    \dext v^2 + 2\Bigl((\mr{D}_b { + 2\mpi_b})\mr{K}^{(n)b}_{a} - (\mr{D}_a + \mr{\pi}_a) \mr{\theta}^{(n)} - 2\mr{F}^{\ b}_{a}B_b \Bigl) \dext \sigma^{a}\dext v \\
    &\quad +\Bigl( d_{\langle ab\rangle} + \f{1}{2}\mr{q}_{ab} ( \mr{K}_c^{(n)d}\mr{K}^{(n)c}_d +2 B_c B^c)\Bigl)\dext \sigma^a \dext \sigma^b \Bigl\} + O(\rho^3),
\end{aligned}
\end{equation}
while the dual Maxwell tensor is
\begin{equation}
\begin{aligned}
\star \bd{F}&=\f{1}{\sqrt{q}}(q_{ac}U^c \pa_\rho A_b + \pa_a A_b)\epsilon^{ab}\ \dext  v\wedge \dext \rho
-\f{1}{\sqrt{q}} q_{ab} \pa_\rho A_c\ \epsilon^{bc}\
\dext  \rho\wedge\dext  \sigma^a\\
&\quad +\f{1}{\sqrt{q}} \Bigl(
(q_{a b}\pa_v A_c
+ q_{a c}\pa_b A_v +2Vq_{ab}\pa_\rho A_c + q_{ad}U^d \pa_c A_b)\ \epsilon^{bc}\\
&\quad +qU^b (U^c\pa_\rho A_c -\pa_\rho A_v )\ \epsilon_{ab}\Bigl) \ \dext v\wedge\dext \sigma^a - \f{\sqrt{q}}{2} (U^c\pa_\rho A_c - \pa_\rho A_v) \epsilon_{ab}\ \dext  \sigma^a\wedge \dext  \sigma^b\\
&=\f{1}{\sqrt{\mr{q}}}\Bigl(\pa_a\mr{A}_b\ \epsilon^{ab}\ \dext  v\wedge\dext \rho - \epsilon^b_{\ a}B_b \ \dext \rho\wedge\dext \sigma^a + \epsilon_a^{\ b}\pa_v \mr{A}_b\ \dext v\wedge\dext\sigma^a \Bigl)+ O(\rho).
\end{aligned}
\label{Fstar}
\end{equation}

\subsubsection{Evolution equations}

In the previous subsection, we used the hypersurface Einstein equations to obtain the radial expansions of the metric's functions. In this subsection, we use the remaining components of the Einstein tensor to derive the evolution equation of the longitudinal expansion, encoded into the $\ell \ell$ component, and of the Hajicek field, encoded into the $\ell a$ component. The leading order of the $\mathbb{E}_{\ell\ell}$ equation yields
\begin{equation}
\begin{aligned}
\mr{\mathbb{E}}_{\ell\ell}=\pa_v \mr{\theta}^{(\ell)} -\mr{\kappa} \mr{\theta}^{(\ell)} + \mr{K}_{(\ell)}^{ab}\mr{K}^{(\ell)}_{ab} +2\mr{q}^{ab}\pa_v \mr{A}_b \pa_v \mr{A}_a,
\end{aligned}
\label{null_ray}
\end{equation}
that is the null Raychaudhuri equation and the leading order of $\mathbb{E}_{\ell a}$ gives
\begin{equation}
\begin{aligned}
\mr{\mathbb{E}}_{\ell a}=(\pa_v +\mr{\theta}^{(\ell)})\mr{\pi}_a -\mr{D}_a (\mr{\kappa}+\mr{\theta}^{(\ell)}) +\mr{D}_b \mr{K}_{a}^{(\ell)b} -2\mr{F}_a^{\ b}\pa_v\mr{A}_{b} ,
\end{aligned}
\label{damour}
\end{equation}
that is the Damour equation. The evolution equation of the transversal extrinsic curvature is encoded into the $\mathbb{E}_{ab}$ component, which reads
\begin{equation}
\begin{aligned}
 \mr{\mathbb{E}}_{ab} &= 2(\pa_v +\mr{\kappa})\mr{K}^{(n)}_{ab} -2\mr{D}_{(a}\mr{\pi}_{b)} -2\mr{\pi}_a\mr{\pi}_b + \mr{\theta}^{(n)} \mr{K}^{(\ell)}_{ab} + \mr{\theta}^{(\ell)} \mr{K}^{(n)}_{ab} - 2\mr{K}^{(\ell)}_{c(a}\mr{K}^{(n)c}_{b)} \\
&\qquad - 2\mr{K}^{(n)}_{c(a}\mr{K}^{(\ell)c}_{b)} + \mr{\cR}_{ab} -2\mr{F}_{ac}\mr{F}^c_{\ b} -4B_{(a}\pa_v\mr{A}_{b)} -\f{1}{2}\mr{q}_{ab}(\mr{R} - |\mr{F}|^2) ,
\end{aligned}
    \label{Rab}
\end{equation}
where $\mr{\cR}_{ab}$ is the Ricci tensor associated with the boundary metric $\mr{q}_{ab}$. Finally, the evolution equation of the Maxwell field comes from the $\mathbb{M}_a$ component, whose leading order reads
\begin{equation}
\begin{aligned}
\mr{\mathbb{M}}_a = 2(\pa_v  +\mr{\mu})B_a -2 \mr{K}^{(\ell)b}_{a} B_b - (\mr{D}_b + 2\mr{\pi}_b)\mr{F}^b_{\ a} +\mtn \pa_v \mr{A}_a -2\mr{K}^{(n)b}_{a}\pa_v \mr{A}_b,
\end{aligned}
\end{equation}
and $\mu = \kappa+ \f{1}{2}\theta^{(\ell)}$ is the \emph{surface tension} of $\cN$. The $\mathbb{M}_v$ component yields
\begin{equation}
    \mr{\mathbb{M}}_v =\mr{D}^a \pa_v \mr{A}_a.
\end{equation}
As shown in \cite{Chandrasekaran:2021hxc}, the Damour and null Raychaudhuri equations can be derived from the conservation law of the null (Carrollian) stress-energy tensor, the latter defined  in \eqref{Tij} via the Weingarten operator.
By decomposing the stress-energy tensor \eqref{Tij} into three contributions, namely 
\begin{equation}
\begin{aligned}
    \cE &= -T^i_{\ j}\ \ell^j n_i \nueq \mr{\theta}^{(\ell)},\\
    \cP_a &= T^i_{\ j}\ n_i\ q^j_{\ a} \nueq \mr{\pi}_a,\\
    \Sigma^{a}_{\ b} &= T^i_{\ j}\ q^j_{\ b} q^a_{\ i}\nueq \mr{K}^{(\ell)a}_{b} -(\mr{\theta}^{(\ell)}+\mr{\kappa})\var_{\ b}^a,\\
\end{aligned}
\end{equation}
which are the energy density $\cE$, the momentum density $\cP_a$ and a spatial stress-energy tensor $\Sigma^a_{\ b}$ respectively, the conservation law $D_iT^i_{\ j}=0$ gives the evolution equation for $\cE$, that is \eqref{null_ray} and for $\cP_a$, namely \eqref{damour}. The evolution equation of the spatial stress-energy tensor $\Sigma^{a}_{\ b}$ should be related to the equation \eqref{Rab}. However, by a simple counting argument, it does not come from the conservation law of the Carrollian tensor. A Carrollian analysis of the evolution equation of the spatial stress-energy tensor $\Sigma^{a}_{\ b}$ will be the subject of future work.

\section{Near-horizon symmetries}\label{sec3}
Now, we commence our Noether analysis for near-horizon geometries. In the previous section, we used the Newman-Unti gauge conditions \eqref{NU} and imposed the near-horizon boundary conditions \eqref{BC}. Therefore, we look for vector field satisfying these gauge and boundary conditions, namely we require
\begin{equation}
    \Lie_\xi g_{v\rho}=0, \qquad \Lie_\xi g_{\rho \rho}=0, \qquad \Lie_\xi g_{\rho a}=0,
\end{equation}
and
\begin{equation}
    \Lie_\xi g_{vv}=O(\rho), \qquad \Lie_\xi g_{\rho a}=O(\rho), \qquad \Lie_\xi g_{ab}=O(1).\label{LieBC}
\end{equation}
The gauge-preserving conditions yield the following vector fields
\begin{equation}
\begin{aligned}
    \xi^v &= \tau(v,x),\\
    \xi^a &= Y^a(v,x) +\pa_b \tau \int d\rho\ g^{ab},\\
    \xi^\rho &= Z(v,x) -\rho\dot{\tau} +\pa_b\tau \int d\rho\ g_{va}g^{ab},
\end{aligned}
\label{diffcom}
\end{equation}
while the boundary conditions \eqref{LieBC} give the following constraints,
\begin{equation}
\pa_v Y^a=0 \qquad \text{and} \qquad Z=0.
\end{equation}
In particular, using the same argument as in \cite{Ciambelli:2021vnn} to obtain the universal corner symmetry algebra, we consider up to the linear order in the $v$-expansion of $\tau$, so we write
\begin{equation}
    \tau = T(x) +vW(x).\label{t+w}
\end{equation}
Therefore, the vector fields generating diffeomorphisms finally read
\begin{equation}
\begin{aligned}
    \xi^v &= T(x)+vW(x)\\
    \xi^a &= Y^a(x) + I^{ab} \pa_b \tau\\
    \xi^\rho &= -\rho W + I^{b}\pa_b\tau,
\end{aligned}
\end{equation}
where
\begin{equation}
    I^{ab} = \int \dext\rho \ g^{ab} \qquad\text{and}\qquad I^b = \int \dext\rho\ g_{va}g^{ab}.
\end{equation}
Next, we want to compute the near horizon symmetry algebra. Demanding that the near horizon symmetry algebra forms a Lie algebra, we impose $\var \tau=\var Y=0$, obtaining
\begin{equation}
    \lim_{\rho\to 0}\ [\![\xi_{(\tau_1, Y_1)}, \xi_{(\tau_2, Y_2)}]\!] = [\bar{\xi}_{(\tau_1, Y_1)}, \bar{\xi}_{(\tau_2, Y_2)}]= \bar{\xi}_{(\tau_{12}, Y_{12})}
\end{equation}
where
\begin{equation}
    \tau_{12}= \tau_1\pa_v \tau_2 + Y^a_1\pa_a \tau_2 -1\leftrightarrow 2, \qquad \text{and}\qquad Y^a_{12} = Y^b_1\pa_b Y^a_2-1\leftrightarrow 2.
\end{equation}
This group is the near-horizon analogue of the Weyl-BMS group found for asymptotically flat spacetimes in \cite{Freidel:2021fxf} (see also \cite{Chandrasekaran:2018aop}), \ie
\begin{equation}
    \fr{g} = \diff(\cS)\loplus \R_v^{\cS},
\end{equation}
which can be rewritten as a double semi-direct sum using \eqref{t+w},
\begin{equation}
     \fr{g} = (\diff(\cS)\loplus \R_W^{\cS})\loplus \R_T^{\cS},
\end{equation}
consisting of $\cS$-diffeomorphisms, super-translations parametrized by $T$ and Weyl super-boosts parametrized by $W$.
Now, the integral terms in \eqref{diffcom} encode the bulk extension of the near-horizon (Carrollian) vector field
\begin{equation}
    \mr{\xi} = \tau \pa_v + Y^a \pa_a,
\end{equation}
and we can evaluate them using the radial expansion of the metric components. Up to the second order in $\rho$, we obtain
\begin{equation}
    I^{ab} = \rho\mr{q}^{ab} - \rho^2 \mr{K}^{ab}_{(n)}+o(\rho^2),\label{Ib}
\end{equation}
and
\begin{equation}
    I^b =  \int \dext \rho \ (\mr{q}^{ab} - 2\rho \mr{K}^{ab}_{(n)})(2\rho \mr{\pi}_a +..) = \rho^2\mr{\pi}^b+o(\rho^2).
\label{IAb}
\end{equation}
Then, plugging \eqref{Ib} and \eqref{IAb} into \eqref{diffcom}, we obtain
\begin{equation}
\begin{aligned}
    \xi^v &= \tau(v,x),\\
    \xi^a &= Y^a(v,x) +\rho( \mr{q}^{ab}\pa_b \tau) - \rho^2\mr{K}_{(n)}^{ab}\pa_b\tau + o(\rho^2) ,\\
    \xi^\rho &= -\rho\dot{\tau} +\rho^2 \mr{\pi}^a\pa_a\tau +o(\rho^2).
\end{aligned}
\label{diffeos}
\end{equation}
In particular, from \eqref{diffeos} we can read off the following vector fields 
\begin{equation}
\begin{aligned}
\xi_T &= T\pa_v + \rho \mr{q}^{ab}\pa_b T \pa_a + \rho^2 \pa_b T\Bigl( \mr{\pi}^b \pa_\rho - \mr{K}_{(n)}^{ab} \pa_a \Bigl)+O(\rho^3),\\
\xi_W &= -\rho W \pa_\rho + v\xi_{T=W}, \\
\xi_Y &=  Y^a \pa_a,\\
\end{aligned}
\end{equation}
generating super-translations, Weyl super-boosts and $\cS$-diffeomorphisms, respectively. \\
Concerning the residual gauge transformations related to the Maxwell field, we need to ensure the radial gauge to be preserved
\begin{equation}
    \var_{(\xi, \varepsilon)} A_\rho = 0,\qquad \ie \qquad \Lie_\xi A_\rho +\pa_\rho \varepsilon=0,
\end{equation}
which yields
\begin{equation}
\begin{aligned}
\varepsilon&= \mr{\varepsilon}(v,\sigma) -\int \dext \rho\ A_a\pa_\rho\xi^a\\
&= \mr{\varepsilon} - \rho \mr{A}^a\pa_a\tau + \f{\rho^2}{2} (B^b - 2\mr{K}^{ab}_{(n)}\mr{A}_a )\pa_b\tau + o(\rho^2) .
\end{aligned}
\end{equation}

\subsubsection*{Internal boost symmetries}
 As shown in \cite{Freidel:2024emv, Ciambelli:2023mir}, there is also a (Carrollian) internal local boost symmetry, associated with the rescaling symmetry
\begin{equation}
    \ell \to e^{\lambda}\ell,
\end{equation}
as also argued in section \ref{nullhyper}. In particular, this internal boost symmetry acts as follows 
\begin{equation}
    \var_\lambda \bd{\ell}=\lambda\bd{\ell}, \qquad \var_\lambda \bd{n}=-\lambda \bd{n}, \qquad \var_\lambda \bd{q}=0.\label{lambda_symm}
\end{equation}
To this symmetry is associated a non-vanishing charge, which is equal to the corner area element \cite{Freidel:2024emv, Ciambelli:2023mir}, as we will show in the following subsections. Moreover, performing an analysis via the Einstein-Cartan formulation of gravity in appendix \ref{EC}, we notice that this symmetry acts exactly as a Lorentz boost, yielding the same charge. Therefore, from now on, we label this internal boost symmetry by $\lambda \nueq \pa_\rho\xi^\rho=-W$.

\subsection{Action on phase space}\label{solspace}
Let us consider a generic metric functional $\cO[\bd{g}]$. The transformation rule of $\cO$ under the near-horizon symmetry group is the following,
\begin{equation}
    \delta_{(\tau, Y)}\cO [\bd{g}] = \int \f{\var \cO}{\var g_{\mu\nu}}\Lie_\xi g_{\mu\nu}.
\end{equation}
Therefore, in this section we analyse the behaviour of the metric functional under symmetry transformations generated by the vector fields \eqref{diffeos}. In particular, we distinguish between two contributions in the transformation of the metric functional 
\begin{equation}
\delta_{(\tau, Y)}\cO [\bd{g}] = \Lie_{\xi} \cO [\bd{g}] +  \Delta_{\xi} \cO [\bd{g}],
\end{equation}
where the first term on the rhs is the homogeneous term, while the last term is an anomaly term. In order to evaluate how the functional transforms under the diffeomorphisms in \eqref{diffeos}, we provide the near-horizon expansion of these vector fields up to the second order in the radial coordinate. For convenience, let us write the vector field as follows
\begin{equation}
\xi_{(\tau,Y)} = \mr{\xi}_{(\tau,Y)}+\rho\xi^{(1)} + \rho^2\xi^{(2)}  + \cdots,
\end{equation}
where
\begin{equation}
    \xi^{(1)} = \mr{q}^{ab}\pa_b\tau \pa_a -\dot{\tau}\pa_\rho,\qquad
    \xi^{(2)} = \mpi^a\pa_a\tau\pa_\rho - \mKnu\pa_b\tau\pa_a,
\end{equation}
and so on. Then, to evaluate the transformation rules of the corner metric $\mr{q}_{ab}$ and the quantity $\lambda_{ab}$, we have to compute the following Lie derivative 
\begin{equation}
\Lie_\xi g_{ab} = \tau \pa_v g_{ab} + \xi^\rho\pa_\rho g_{ab} + \xi^c\pa_c g_{ab} + 2g_{c(b}\pa_{a)}\xi^c + 2g_{v(b}\pa_{a)}\xi^v 
\end{equation}
up to the first order in $\rho$. We obtain
\begin{equation}
\begin{aligned}
\var_{(\tau, Y)}\mr{q}_{ab} &=(\tau \pa_v +\Lie_Y) \mr{q}_{ab},\\
\var_{(\tau, Y)} \lambda_{ab} &=(\tau\pa_v +\Lie_Y -\dot{\tau}) \lambda_{ab} +2 \mr{D}_{(a}\mr{D}_{b)}\tau + 4\mr{\pi}_{(b}\mr{D}_{a)}\tau,
\end{aligned}
\label{var_CAb}
\end{equation}
and the leading term of the volume form yields
\begin{equation}
\var_{(\tau, Y)} \sqrt{\mr{q}} = (\tau \pa_v +\mr{D}_aY^a) \sqrt{\mr{q}}.\label{varsqrtq}
\end{equation}
For our purpose, we are interested in the transformations properties of the transversal and longitudinal expansion and shear, which read
\begin{equation}
\begin{aligned}
\var_{(\tau, Y)}\mr{\sigma}^{(n)}_{ab} &=(\tau\pa_v +\Lie_Y -\dot{\tau}) \mr{\sigma}^{(n)}_{ab} + \mr{D}_{\langle a}\mr{D}_{b\rangle}\tau + 2\mr{\pi}_{\langle b}\mr{D}_{a\rangle}\tau,\\
\var_{(\tau, Y)}\mr{\sigma}^{(\ell)}_{ab} &=(\tau\pa_v +\Lie_Y +\dot{\tau}) \mr{\sigma}^{(\ell)}_{ab} ,\\
\var_{(\tau, Y)}\mr{\theta}^{(n)} &=(\tau\pa_v +\Lie_Y -\dot{\tau}) \mr{\theta}^{(n)} + \mr{D}^2\tau + 2\mr{\pi}^a\mr{D}_{a}\tau,\\
\var_{(\tau, Y)}\mr{\theta}^{(\ell)} &=(\tau\pa_v +\Lie_Y +\dot{\tau}) \mr{\theta}^{(\ell)}.
\end{aligned}
\label{tr_rule1}
\end{equation}
Then, by evaluating the Lie derivative of the $g_{vv}$ and $g_{va}$ components, we obtain
\begin{equation}
\begin{aligned}
\var_{(\tau, Y)} \mr{\kappa} &= (\tau\pa_v +\Lie_Y+\dot{\tau} )\mr{\kappa},\\
\var_{(\tau, Y)} \mr{\pi}_a &=(\tau \pa_v +\Lie_Y  )\mr{\pi}_a + \mr{\kappa}\pa_a\tau +\pa_a \dot{\tau} - \mr{K}^{(\ell)b}_a\pa_b\tau.
\end{aligned}
\label{tr_rule2}
\end{equation}
Taking into account the contribution of the large gauge transformations, the components of the Maxwell field transforms according to the following rule
\begin{equation}
\begin{aligned}
    \var_{(\xi, \varepsilon)} A_a &= \xi^\mu\pa_\mu A_a + A_\mu\pa_a\xi^\mu + \pa_a\varepsilon,
\end{aligned}
\end{equation}
which explicitly yields
\begin{equation}
\begin{aligned}
\var_{(\xi, \varepsilon)} \mr{A}_a & = (\tau\pa_v + \Lie_Y )\mr{A}_a  + \pa_a \mr{\varepsilon},\\
\var_{(\xi, \varepsilon)} B_a &= (\tau\pa_v +\Lie_Y -\dot{\tau}) B_a + \mr{F}^b_{\  a}\pa_b\tau ,
\end{aligned}
\end{equation}
up to the linear order in the radial coordinate.

\subsection{Pre-symplectic potential}
Now, having characterised the transformation rules of the various metric's functionals, we can proceed our treatment by applying the formalism outlined in section \ref{CPS}. The pre-symplectic potential  current of  EM gravity reads as follows
\begin{equation}
    \theta^\mu_{\EM}[g, A;\var g, \var A] = \f{1}{2}(g^{\alpha\beta} \delta\Gamma^\mu_{\alpha\beta}-g^{\alpha\mu} \delta \Gamma^\beta_{\alpha\beta}- 4F^{\mu\nu}\var A_\nu).
\end{equation}
The charge aspect is
\begin{equation}
     \dext\bd{q}^\EM_\xi = I_\xi \bd{\theta}_{\EM} - (\iota_\xi \bold{L} + \bd{a}_\xi)_{\EM}\label{78}
\end{equation}
and therefore the charge is obtained by integrating the charge aspect in \eqref{78} on the corner, \ie
\begin{equation}
    \cQ^{\EM}_\xi = \int_\cS \bd{q}^\EM_\xi.
\end{equation}
In order to compute the charge and the flux content of the theory, we need to evaluate the component of the pre-symplectic potential (see appendix \ref{pre-pot_der}). The pullback of the pre-symplectic potential on $\cN$ yields
\begin{equation}
\begin{aligned}
    \Theta^{\cN}_{\EM}[\bd{g}, \var \bd{g}]&= \int_\cN \theta_\EM^\mu[\bd{g}, \var \bd{g}]\ \ell_\mu \ \eps_\cN\\
    &=-\f{1}{2} \int_\cN \Bigl[ \Bigl(\sigma^{ab}_{(\ell)} -\mu q^{ab}\Bigl)\var q_{ab} +2\pi_a \var U^a +4\ell_\mu F^{\mu\nu}\var A_\nu \Bigl]\ \eps_\cN \\
    &\quad -\var(\int_\cN(\kappa +\theta^{(\ell)})\ \eps_\cN) -\f{1}{2}\int_{\cN} D\cdot \var U\ \eps_{\cN}.
\end{aligned}
\label{Npot}
\end{equation}
The expression in \eqref{Npot}  is consistent with the results found via the Carrollian approach in  \cite{Freidel:2024emv}\cite{Ciambelli:2023mir} and in \cite{Chandrasekaran:2021hxc}.
From the ambiguity of the pre-symplectic potential, \ie
\begin{equation}
    \bd{\theta} \to \bd{\theta}'=\bd{\theta} + \var\bd{\ell}_b -\dext\bd{\vartheta} ,\label{amb_pot}
\end{equation}
we can define a boundary Lagrangian from the total field-variation term in \eqref{Npot}, which reads
\begin{equation}
    \bd{\ell}_\cB = -\Bigl(\kappa+\theta^{(\ell)}\Bigl)\ \eps_\cN,\label{boundlagr}
\end{equation}
and a corner potential
\begin{equation}
    \bd{\vartheta}= \f{1}{2}\var U^a\  \iota_a \eps_\cN.
\end{equation}
Therefore, once the boundary Lagrangian and the corner potential have been identified, the Einstein-Maxwell pre-symplectic potential \eqref{Npot} can be written as
\begin{equation}
    \bd{\theta}_\EM = \bd{\theta}^c +\var \bd{\ell}_\cB - \dext\bd{\vartheta}\label{EH&c}
\end{equation}
by means \eqref{amb_pot}, where $\bd{\theta}^c$ represents the canonical pre-symplectic potential \cite{Ciambelli:2023mir, Freidel:2024emv} and its integral on $\cN$ reads as follows
\begin{equation}
\Theta^{c}[\bd{g}, \var\bd{g}] := -\f{1}{2} \int_\cN \Bigl[\Bigl(\sigma_{ab}^{(\ell)} -\mu q_{ab}\Bigl)\var q^{ab} + 2\pi_a \var U^a + 4\ell_\mu F^{\mu\nu}\var A_\nu \Bigl]\ \eps_\cN.   \label{thetac}
\end{equation}
Using again the pre-symplectic potential ambiguity \eqref{EH&c} and the formula in \eqref{mod_Q}, the difference between the Einstein-Maxwell charge and canonical charge reads as follows \cite{Freidel:2021cjp},
\begin{equation}
\cQ^\EM_\xi - \cQ^c_\xi = \int_S (\iota_\xi\bd{\ell}_\cB -I_\xi \bd{\vartheta}),
\label{mod_L}
\end{equation}
where $\cQ^c$ is the Noether charge associated with the canonical pre-symplectic potential $\bd{\theta}^c$, and the canonical charge aspect reads
\begin{equation}
    \dext\bd{q}^c_\xi = I_\xi\bd{\theta}^c -\iota_\xi(\bold{L}_\EM-\dext\bd{\ell}_\cB) -\bd{a}^\EM_\xi +\Delta_\xi\bd{\ell}_\cB, \label{qc}
\end{equation}
where we used \eqref{EH&c}, \eqref{anom_def} and the Cartan's magic formula.

\subsection{Noetherian charges and fluxes}
In this section, we want to compute the canonical Noether charges associated with the diffeomorphisms in \eqref{diffeos}. The Einstein-Maxwell charges follow straightforwardly from \eqref{mod_L}. From \eqref{qc} the canonical Noether charge associated with the diffeomorphisms in \eqref{diffcom} reads as follows
\begin{equation}
    \cQ^c_\xi = I_\xi \Theta^c + \int_\cN (\iota_\xi \dext \bd{\ell}_\cB+\Delta_\xi \bd{\ell}_\cB) -\int_\cN (\iota_\xi\bold{L}_\EM+\bd{a}^\EM_\xi ) .\label{chargecan}
\end{equation}
In this work we are interested in computing only the leading-order charges on the horizon. Firstly, let us compute the anomaly of the boundary Lagrangian in \eqref{boundlagr}.
The anomaly of the boundary Lagrangian comes from the following formula
\begin{equation}
    (\var_\xi-\Lie_\xi)\bd{\ell}_\cB = -(\var_\xi-\Lie_\xi)\Bigl(\kappa+\theta^{(\ell)}\Bigl)\ \eps_\cN.\label{blanom}
\end{equation}
In particular, we are interested in evaluating the expression in \eqref{blanom} at leading order. Using the formulas derived in the subsection \ref{solspace}, one can readily see that the anomaly of the boundary Lagrangian is zero at the leading level. Moreover, the term
\begin{equation}
    \iota_\xi \dext\bd{\ell}_\cB = -\xi^\rho \pa_\rho\Bigl(\kappa+\theta^{(\ell)}\Bigl)\ \eps_\cN
\end{equation}
yields a sub-leading contribution to the charge and additionally the last integral in \eqref{chargecan} vanishes on-shell at the leading order. Now, having exposed the previous reasons, let us compute the canonical charge associated with diffeomorphisms generated by the vector fields in \eqref{diffcom}. Then, the leading order of the near-horizon charge associated with super-translations is
\begin{equation}
\begin{aligned}
    \cQ^c [\xi_T] &= \Theta^\cN_{c}[\bd{g}, \var_T\bd{g}] + \Theta^\cN_{\mathrm{M}}[\bd{A}, \var_T\bd{A}]\\
    &= -\f{1}{2} \int_\cN \Bigl[\Bigl(\mr{K}_{ab}^{(\ell)} -(\mr{\kappa} + \mr{\theta}^{(\ell)}) \mr{q}_{ab}\Bigl)\var_T \mr{q}^{ab} - 4 \mr{F}^{\rho c}\var_T \mr{A}_c \Bigl]\ \mr{\eps}_\cN\\
    &= - \int_\cN T\Bigl(\mr{K}_{ab}^{(\ell)}\mr{K}^{ab}_{(\ell)} - \mr{\kappa}\mr{\theta}^{(\ell)} +\pa_v\mr{\theta}^{(\ell)} + 2\mr{q}^{ab}\pa_v\mr{A}_b\pa_v\mr{A}_a\Bigl) \mr{\eps}_\cN \\
    &\quad+ \int_\cS T\mr{\theta}^{(\ell)} \sqrt{\mr{q}}\ \dext^2\sigma,\\
\end{aligned}
\end{equation}
where we used the transformation rules in \eqref{tr_rule1}-\eqref{tr_rule2}, and the following relation $\pa_v(\mthell \sqrt{\mr{q}}) = ((\mthell)^2 +\pa_v\mthell)\sqrt{\mr{q}}$. The charge associated with diffeomorphisms of $\cS$ is
\begin{equation}
\begin{aligned}
    \cQ^c [\xi_Y] &= \Theta^\cN_{c}[\bd{g}, \var_Y\bd{g}] + \Theta^\cN_{\mathrm{M}}[\bd{A}, \var_Y\bd{A}]\\
    &= -\f{1}{2} \int_\cN \Bigl[\Bigl(\mr{K}_{ab}^{(\ell)} -(\mr{\kappa} + \mr{\theta}^{(\ell)}) \mr{q}_{ab}\Bigl)\var_Y \mr{q}^{ab} - 4 \mr{F}^{\rho c}\var_Y \mr{A}_c \Bigl]\ \mr{\eps}_\cN\\
    &= \int_\cN Y^a \Bigl(\mr{D}_b \mr{K}^{(\ell)b}_{a} - \mr{D}_a (\mr{\kappa}+\theta^{(\ell)}) +\pa_v \mr{\pi}_a +\mr{\pi}_a \mr{\theta}^{(\ell)} - 2\mr{F}^{\ b}_a\pa_v\mr{A}_b\Bigl) \mr{\eps}_\cN\\
    &\quad + \int_\cN Y^a \mr{A}_a \mr{\mathbb{M}}_v\ \mr{\eps}_\cN - \int_\cS Y^a \mr{\pi}_a\ \sqrt{\mr{q}}\ \dext^2\sigma,
\end{aligned}
\end{equation}
where we again used the transformation rules in \eqref{tr_rule1}-\eqref{tr_rule2}, and the relations in \eqref{D_rel}. The leading charge associated with $\xi_W$ is $\cQ^c_{\xi_W} = v\cQ^c_{\xi_{T=W}}$. These are the well-known Carrollian charges which stem by integrating the Brown-York charge density $\bd{j}_\xi = -T^j_{\ i} \xi^i \eps_j$ on $\cN$ \cite{Chandrasekaran:2021hxc, Freidel:2024emv, Ciambelli:2023mir}.\\
As argued in the previous section, there is also an internal symmetry parametrized by $\lambda$. 
From \eqref{lambda_symm} it is straightforward to see that $I_\lambda \Theta^c=0$ at the leading order, but the anomaly of the boundary Lagrangian is non-vanishing and is the only contribution to the charge associated with $\lambda$. In particular,
\begin{equation}
\begin{aligned}
    \Delta_\lambda \theta^{(\ell)}= \lambda \theta^{(\ell)},\qquad \Delta_\lambda \kappa = \ell^\mu\pa_\mu \lambda +\lambda\kappa, \qquad \Delta_\lambda \eps_\cN = -\lambda \eps_\cN,
\end{aligned}
\end{equation}
and therefore the Noether charge associated with local boosts is
\begin{equation}
    \cQ^c_\lambda = \int_\cN \ell^\mu \pa_\mu \lambda \ \eps_\cN = -\int_\cS W\ \sqrt{\mr{q}}\ \dext^2\sigma.
\end{equation}
In summary, the near-horizon canonical charges are
\begin{equation}
\begin{aligned}
    \cQ^c_{\xi_T} &= \int_\cS T\mr{\theta}^{(\ell)}\sqrt{\mr{q}}\ \dext^2\sigma,\qquad
    &\cQ^c_{\xi_Y} &= -\int_\cS Y^a\mr{\pi}_a \sqrt{\mr{q}}\ \dext^2\sigma,\\
    \cQ^c_{\xi_W} &= v \cQ^c_{\xi_{T=W}},\qquad
    &\cQ^c_{\lambda} &= -\int_\cS W\ \sqrt{\mr{q}}\ \dext^2\sigma.
\end{aligned}
\end{equation}
Now, by means the \eqref{mod_L}, the Einstein-Maxwell charges are
\begin{equation}
\cQ_T^\EM = -\int_\cS T \mr{\kappa} \sqrt{\mr{q}}\ \dext^2\sigma,\qquad 
\cQ_W^\EM = v\cQ^\EM_{T=W} ,\qquad \cQ_Y^\EM = \cQ_Y^c,
\end{equation}
and $\cQ_\lambda^\EM = \cQ_\lambda^c$, which coincide with the Einstein-Cartan charges found in appendix \ref{EC}. 
In particular, the combination of the Weyl super-boost and the local boost yields the following Noether charge
\begin{equation}
    \cQ^\EM_{(\xi_W, \lambda)} = -\int_\cS W\Bigl( 1 - v\pa_v \Bigl) \sqrt{\mr{q}}\ \dext^2\sigma,
\end{equation}
which is exactly the dynamical entropy discussed in \cite{Rignon-Bret:2023fjq, Hollands:2024vbe, Visser:2024pwz}. In other words, we claim that the Weyl super-boost charge encodes the dynamical corrections to the Bekenstein-Hawking (BH) entropy formula \cite{Bekenstein:1973ur, Hawking:1975vcx}. Hence, the dynamical entropy is the Noether charge associated with a symmetry transformation consisting of Weyl super-boosts and local boosts. Although a more detailed investigation of the connection between the Weyl super-boost and the dynamical corrections to the BH formula is an interesting topic to pursue, it goes beyond the scope of this work and will be discussed in the future.\\
The general expression for the electric charge can be easily computed as follows
\begin{equation}
\begin{aligned}
    \cQ^e_\M[\varepsilon] \heq \f{1}{2}\int_{\cS} \varepsilon \star F_{ab} \ \dext \sigma^a\wedge\dext \sigma^b ,
\end{aligned} 
\end{equation}
and from \eqref{Fstar} yields a vanishing contribution at the leading order. The dual (or magnetic) charge is
\begin{equation}
\begin{aligned}
    \cQ^m_\M[\varepsilon] &\heq \int_{\pa\cN} \mr{\varepsilon}\ \epsilon^{ab} \pa_a \mr{A}_b\ \sqrt{\mr{q}}\ \dext^2\sigma.
\end{aligned} 
\end{equation}

\vspace{0.2cm}

\noindent
In order to compute the charge algebra from \eqref{Q_algebra}, we need to evaluate the flux content of the system. Here, we provide a derivation of the Noetherian fluxes associated with the vector fields in \eqref{diffeos} using the following formula
\begin{equation}
\cF_\xi^\EM := \cF_\xi^{\theta^\EM} + \cF^\EM_{\var\xi} =\int_S (\iota_\xi\bd{\theta}^\EM +\bd{A}^\EM_\xi) + \int_S \bd{q}^\EM_{\delta\xi}.
\label{flux}
\end{equation}
However, to be coherent with the previous results, we have to provide a derivation of the canonical Noetherian flux. From \eqref{EH&c}, we have
\begin{equation}
\begin{aligned}
\iota_\xi \bd{\theta}^\EM &= \iota_\xi \bd{\theta}^c +\iota_\xi\var \bd{\ell}_\cB -\iota_\xi \dext \bd{\vartheta}\\
&= \iota_\xi \bd{\theta}^c -\iota_{\var\xi} \bd{\ell}_\cB +\var \iota_\xi \bd{\ell}_\cB -\Lie_\xi\bd{\vartheta} + \dext \iota_\xi \bd{\vartheta},
\end{aligned}
\end{equation}
and
\begin{equation}
\begin{aligned}
  \dext\bd{A}^\EM_\xi &= \Delta_\xi \bd{\theta}^\EM - \var \bd{a}_\xi^\EM + \bd{a}_{\var\xi}^\EM\\
  &= \Delta_\xi \bd{\theta}^c + \Delta_\xi \var \bd{\ell}_\cB -\Delta_\xi \dext\bd{\vartheta}  - \var \bd{a}_\xi^\EM + \bd{a}_{\var\xi}^\EM\\
  &= \Delta_\xi \bd{\theta}^c +\var \Delta_\xi \bd{\ell}_\cB - \Delta_{\var\xi} \bd{\ell}_\cB  
  -\dext\Delta_\xi\bd{\vartheta}  - \var \bd{a}_\xi^\EM + \bd{a}_{\var\xi}^\EM\\
  &= \Delta_\xi \bd{\theta}^c -\dext(\var_\xi-\Lie_\xi -I_{\var\xi}) \bd{\vartheta}  - \var \bd{a}_\xi^c + \bd{a}_{\var\xi}^c,
\end{aligned}
\end{equation}
so that
\begin{equation}
    \dext \bd{A}^c_\xi =  \dext (\bd{A}^\EM_\xi + \Delta_\xi\bd{\vartheta})= \Delta_\xi\bd{\theta}^c - \var \bd{a}_\xi^c + \bd{a}_{\var\xi}^c,
\end{equation}
where $\bd{a}^c_\xi = \bd{a}^\EM_\xi -\Delta_\xi \bd{\ell}_\cB$. Therefore, from \eqref{mod_L} we have
\begin{equation}
\bd{q}^\EM_{\var\xi} =\bd{q}^c_{\var\xi}+ \iota_{\var\xi}\bd{\ell}_\cB -I_{\var\xi} \bd{\vartheta}.
\end{equation}
The canonical Noetherian flux reads as follows 
\begin{equation}
    \cF^c_\xi = \int_\cS (\iota_\xi\bd{\theta}^c + \bd{A}^c_\xi +\bd{q}_{\var\xi}^c) ,
\end{equation}
and we recover the formula in \eqref{mod_Q} for which
\begin{equation}
    \cF_\xi^\EM - \cF_\xi^c = \int_\cS (\var \iota_\xi \bd{\ell}_\cB - \var_\xi \bd{\vartheta}).\label{FEH&C}
\end{equation}
Although the Lagrangian and symplectic anomaly vanishes in the Einstein-Maxwell formulations, in the canonical formulation it may not because of the terms $\Delta_\xi \bd{\ell}_\cB$ and $\Delta_\xi\bd{\vartheta}$. However, by taking a look at the transformation rules in subsection \ref{solspace} under the diffeomorphisms in \eqref{diffeos}, at the leading level we have
\begin{equation}
    \Delta_\xi \mr{\bd{\ell}}_\cB =0 \qquad \text{and} \qquad \Delta_\xi\mr{\bd{\vartheta}}=0.
\end{equation}
Using the expression in \eqref{symanom} and plug it into the expression of the anomaly of the pre-symplectic potential, it follows that the symplectic anomaly and the $\bd{q}_{\var\xi}$-term give vanishing contributions at the leading order. Therefore, the EM Noetherian flux simply reads as the symplectic flux,
\begin{equation}
    \int_\cS\iota_\xi \bd{\theta}^\EM = \int_\cS d^2 \sigma \ \sqrt{q}\ (\xi^\rho \theta^v-\xi^v \theta^\rho).\label{iotathetaEH}
\end{equation}
Now, using the expressions furnished in appendix \ref{pre-pot_der}, we have the following contributions to the integrand in \eqref{iotathetaEH}
\begin{equation}
\begin{aligned}
    \xi^v \theta^\rho &= \tau \Bigl(\var \mr{\kappa} -\f{1}{2}\mr{K}^{(\ell)}_{ab} \var \mr{q}^{ab} +\var \mr{\theta}^{(\ell)} \Bigl) +O(\rho)
\end{aligned}
\end{equation}
and
\begin{equation}
\begin{aligned}
    \xi^\rho \theta^v = \rho\dot{\tau} K^{(n)}_{ab}\var \mr{q}^{ab}+O(\rho^2).
\end{aligned}
\end{equation}
Hence, we have
\begin{equation}
\begin{aligned}
    \cF^\EM_{\xi_T}=- \int_\cS \dext^2\sigma\ \sqrt{\mr{q}}\ T\Bigl(\delta (\mr{\kappa}+\mr{\theta}^{(\ell)}) -\f{1}{2}\mr{K}^{(\ell)}_{ab} \var \mr{q}^{ab} +2 \mr{q}^{ab}\pa_v \mr{A}_b\var \mr{A}_a \Bigl),
\end{aligned}
\label{FEHT}
\end{equation}
while $\cF^\EM_{\xi_W}=\cF^\EM_{v\xi_{T=W}}$ and $\cF^\EM_{\xi_Y}=0$. The canonical Noetherian flux comes straightforwardly from the \eqref{FEH&C}, yielding the following contribution
\begin{equation}
\begin{aligned}
    \cF^c_{\xi_T}= \f{1}{2}\int_\cS \dext^2\sigma\ \sqrt{\mr{q}}\ T\Bigl[ \Bigl(\mr{K}^{(\ell)}_{ab} -(\mr{\kappa} +\mr{\theta}^{(\ell)})\mr{q}_{ab} \Bigl) \var \mr{q}^{ab} -4 \mr{q}^{ab}\pa_v \mr{A}_b\var \mr{A}_a\Bigl],
\end{aligned}
\label{FcT}
\end{equation}
and again $\cF^c_{\xi_W}=\cF^c_{v\xi_{T=W}}$ and $\cF^c_{\xi_Y}=0$. Finally, since $\var W=0$, we also have $\cF^\EM_\lambda=\cF^c_\lambda=0$. As expected, the electromagnetic flux vanishes at the leading order, \ie
\begin{equation}
    \cF^\M[\varepsilon] = \int_\cS \bd{q}_{\var\varepsilon}\heq 0.
\end{equation}

\subsection{Charge algebra and dynamics}\label{alg_fb}
In this final section, we provide a derivation of the near-horizon charge algebra through the generalized Barnich-Troessaert bracket \cite{Freidel:2021cjp}
\begin{equation}
    \{\cQ_\xi, \cQ_\zeta\}_L \heq  \var_\xi\cQ_\zeta - I_\zeta \cF_\xi + \cK_{(\xi, \zeta)},\label{ch_alg}
\end{equation}
and we show that (some of) the Einstein's equation follows from the so called \emph{flux-balance law},
\begin{equation}
    \var_\xi \cQ_\zeta - I_\zeta \cF_\xi + \cK_{(\xi, \zeta)} + \cQ_{[\![\xi, \zeta]\!]} = \int_\cS \iota_\xi \bd{C}_\zeta,
    \label{fl_law}
\end{equation}
where the rhs of \eqref{fl_law} represents the constraint
\begin{equation}
    \bd{C}_\zeta = \zeta^\mu C_{\mu}^{\ \nu}\eps_\nu,
\end{equation}
which vanishes on-shell. In particular, the structure of \eqref{fl_law} is invariant under a Lagrangian shift \cite{Freidel:2021cjp} and therefore we derive the Damour and null Raychaudhuri equations by using the Einstein-Maxwell Noether charges and fluxes. Then, let us begin by computing the charge algebra associated with (near-horizon) super-translations. We have
\begin{equation}
\begin{aligned}
\var_{\xi_{T_2}}\cQ^\EM_{\xi_{T_1}}&=-\int_\cS T_1 T_2 \Bigl(\pa_v \mr{\kappa}  + \mr{\kappa} \mthell\Bigl) \sqrt{\mr{q}} \ \dext^2\sigma\\
\end{aligned}
\end{equation}
and
\begin{equation}
\begin{aligned}
I_{\xi_{T_1}}\cF^\EM_{\xi_{T_2}} &=  -\int_\cS  T_1T_2\Bigl(\pa_v \mr{\kappa}+ \pa_v \mr{\theta}^{(\ell)} -\mr{K}^{(\ell)}_{ab} \mr{K}_{(\ell)}^{ab} + 2\mr{q}^{ab}\pa_v\mr{A}_b\pa_v\mr{A}_a \Bigl)\sqrt{\mr{q}} \ \dext^2\sigma.
\end{aligned}
\end{equation}
Putting the above contributions together, we have the null Raychaudhuri equation 
\begin{equation}
\begin{aligned}
    \{\cQ_{\xi_{T_1}}, \cQ_{\xi_{T_2}}\}_\EM &= -\int_\cS  T_1T_2\Bigl( (\pa_v -\mr{\kappa})\mr{\theta}^{(\ell)} +\mr{K}^{(\ell)}_{ab} \mr{K}_{(\ell)}^{ab} + 2\mr{q}^{ab}\pa_v\mr{A}_b\pa_v\mr{A}_a \Bigl)\sqrt{\mr{q}} \ \dext^2\sigma,\\
\end{aligned}
\end{equation}
indeed $\cQ^\EM_{[\![T_1, T_2]\!]}\heq 0$. The Damour equation can be recovered by evaluating the \eqref{fl_law} for $\xi = \hat{v}:=\pa_v$ and $\zeta = \xi_Y$. We have that
\begin{equation}
\begin{aligned}
\var_{\hat{v}}\cQ^\EM_{\xi_Y} &= -\int_\cS Y^a (\pa_v +\mr{\theta}^{(\ell)}) \mr{\pi}_a \ \sqrt{\mr{q}}\ \dext^2\sigma ,
\end{aligned}
\end{equation}
and
\begin{equation}
\begin{aligned}
-I_{\xi_Y}\cF^\EM_{\hat{v}} &= \int_\cS Y^a\Bigl(\mr{D}_a(\mr{\kappa} +\mthell) -\mr{D}_b\mr{K}^{(\ell)b}_{a} +2\mr{F}_a^{\ b}\pa_v\mr{A}_b\Bigl) \ \sqrt{\mr{q}} \ \dext^2\sigma\\
&\quad +2\int_{\cS} Y^a\mr{A}_a \mr{D}^b\pa_v\mr{A}_b\ \sqrt{\mr{q}} \ \dext^2\sigma. 
\end{aligned}
\end{equation}
Then, we obtain
\begin{equation}
\begin{aligned}
\var_{\hat{v}} \cQ^\EM_{\xi_Y} - I_{\xi_Y} \cF^\EM_{\hat{v}} &= -\int_\cS Y^a \Bigl( (\pa_v + \mr{\theta}^{(\ell)})\mr{\pi}_a - \mr{D}_a \mu + \mr{D}_b \sigma_{\ a}^{(\ell)b} - 2\mr{F}^{\ b}_a\pa_v\mr{A}_b\Bigl) \sqrt{\mr{q}}\ \dext^2\sigma\\
&\quad  +2\int_{\cS} Y^a\mr{A}_a \mr{D}^b\pa_v\mr{A}_b\ \sqrt{\mr{q}} \ \dext^2\sigma,
\end{aligned}
\end{equation}
where $-Q^\EM_{[\![\hat{v},\xi_Y]\!]}=0$ and the last line is the Maxwell equation $\mr{\mathbb{M}}_v$. In particular, we also obtain
\begin{equation}
    \cQ_{[\![\xi_Y, \xi_T]\!]} \heq -\var_{\xi_Y} \cQ_{\xi_T} = \int_\cS (Y^a\pa_a T) \mr{\kappa}\ \sqrt{\mr{q}}\ \dext^2\sigma.
\end{equation}
The other brackets follow straightforwardly, and we obtain 
\begin{equation}
\begin{aligned}
\var_{\xi_{Y_2}}\cQ^\EM_{\xi_{Y_1}} &= \int_\cS (\mr{\pi}_b Y^a_1\mr{D}_a Y^b_2 -\mr{\pi}_a Y^b_2\mr{D}_b Y^a_1)\sqrt{\mr{q}}\ \dext^2\sigma \\
&= -\cQ^\EM_{[\![\xi_{Y_2}, \xi_{Y_1}]\!]},
\end{aligned}
\end{equation}
since $\cF^\EM_{\xi_Y}=0$, and 
\begin{equation}
\begin{aligned}
&\var_{\xi_T} \cQ_\lambda^\EM -I_\lambda \cF_{\xi_T}^\EM = \int_\cS T \lambda \mr{\kappa} \sqrt{\mr{q}}\ \dext^2\sigma,\qquad \var_{\xi_Y}\cQ_\lambda^\EM = \int_\cS Y^a  \pa_a \lambda \sqrt{\mr{q}}\ \dext^2\sigma.
\end{aligned}
\end{equation}
Summarizing the above results, the near-horizon charge algebra is the following
\begin{equation}
\begin{aligned}
\{\cQ_{\xi_{T_1}}, \cQ_{\xi_{T_2}}\}^\EM &\heq 0,\qquad& \{\cQ_{\xi_T}, \cQ_{\lambda}\}^\EM &\heq -\cQ_{\xi_{T= T\lambda}},\\
\{\cQ_{\xi_T}, \cQ_{\xi_Y}\}^\EM &\heq -\cQ^\EM_{\xi_{T=Y^a\pa_a T}},\qquad& \{\cQ_{\lambda_1}, \cQ_{\lambda_2}\}^\EM &\heq 0,\\
\{\cQ_{\xi_{Y_1}}, \cQ_{\xi_{Y_2}}\}^\EM &\heq -\cQ^\EM_{\xi_{[Y_1, Y_2]}}, \qquad& \{\cQ_{\xi_Y}, \cQ_{\lambda}\}^\EM &\heq -\cQ^\EM_{\lambda=Y^a\pa_a \lambda},\\
\end{aligned}
\label{algebra_NH}
\end{equation}
and the commutation relations for $\cQ^\EM_{\xi_W}$ follows by substituting $\cQ^\EM_{\xi_T=vW}$.

\section{Conclusion}
In this work, we have conducted the analysis of the near-horizon symmetries of a four-dimensional non-extremal black hole in the Einstein-Maxwell theory. The study of the corner symmetry algebra has been carried out by invoking the Noetherian split for charges and fluxes introduced in \cite{Freidel:2021cjp}. As emphasized in \cite{Freidel:2021cjp}, we also outlined the importance of the inclusion of the anomaly operator in the covariant phase space formalism, used in the computation of the local boost charge in the canonical formulation. Moreover, in the same spirit of \cite{Freidel:2021fxf}, we demonstrate that  demanding that the generalized Barnich-Troassert bracket \eqref{ch_alg} gives a representation of the symmetry algebra, the null Raychaudhuri and Damour equations, and the $v$-component of the Maxwell equations emerge holographically on the corner.\\
In particular, our results suggest that the (electric) large gauge transformations act trivially on the horizon, since the electric charge is identically zero. This conclusion is reached due to the boundary conditions used in this work, which kill the leading and sub-leading $v$-component of the gauge field, as can be seen in radial Maxwell equation.\\
Additionally, by also providing a formulation of the problem in the tetrad formalism, we highlighted a connection between the Carrollian internal boost charge and the Lorentz boost charge, whose value is equal to the corner area element. In \cite{Ciambelli:2023mir, Shajiee:2025cxl}, it was shown that this charge provides a notion of gravitational entropy and could represent a good candidate to describe generalized entropy. Furthermore, and perhaps more interestingly, we showed that the Weyl super-boosts encode the dynamical corrections to the Bekenstein–Hawking entropy formula. Indeed, the Weyl super-boosts and the local boosts must be grouped together, and their associated Noether charge is exactly the dynamical entropy discussed in \cite{Rignon-Bret:2023fjq, Hollands:2024vbe, Visser:2024pwz}. However, a detailed analysis of the dynamical entropy goes beyond the goal of this paper and we hope that the connection highlighted in this work could be helpful for future investigations.\\
To conclude, our work currently lacks a derivation of the spacelike Einstein equations $\mathbb{E}_{\langle ab\rangle} =0$ through symmetry considerations. Achieving such a derivation would suggest the existence of a spin-2 symmetry generator, thereby implying an enlargement of the gravitational symmetry group in the near-horizon region.

\appendix
\section{Derivation of pre-symplectic potential}\label{pre-pot_der}
In this appendix, we outline the derivation of the Einstein-Hilbert pre-symplectic potential in \eqref{Npot} using the formula 
\begin{equation}
\begin{aligned}
    \theta^\mu_\EH [\bd{g}, \var \bd{g}]&=\f{1}{2}(g^{\mu\sigma} \nabla^\nu \delta g_{\nu\sigma} -\nabla^\mu \delta g)\\
    &=\f{1}{2}(g^{\alpha\beta} \delta \Gamma^\mu_{\alpha\beta}-g^{\alpha\mu} \delta \Gamma^\beta_{\alpha\beta}).
\end{aligned}
\label{ehcurr}
\end{equation}
In particular, we are interested in computing the pull-back of the pre-symplectic potential on the null hypersurface $\cN$, which is
\begin{equation}
    \bd{\theta}_\EH^\cN = \theta^\mu_\EH \ell_\mu \ \eps_\cN.
    \label{141}
\end{equation}
The components of \eqref{ehcurr} we need to evaluate are the $\rho-$ and $v-$ components. Using the list of Christoffel symbols furnished in appendix \ref{chrsymb}, we obtain
\begin{equation}
\begin{aligned}
\theta_{\EH}^\rho [\bd{g}, \var \bd{g}]&=\f{1}{2}\Bigl(g^{\alpha\beta} \delta \Gamma^\rho_{\alpha\beta} -g^{\alpha\rho} \delta\Gamma^\beta_{\alpha\beta}\Bigl) \\
&=\f{1}{2}\Bigl[2\var\kappa +2\pi_a\var U^a +q^{ab}\var K^{(\ell)}_{ab} +2V\var \theta^{(n)}+\f{1}{2}\var(q^{ab}\pa_v q_{ab})\\
&\quad + q^{ab}\var(V K^{(n)}_{ab}) + D_a\var U^a -\var(D_a U^a)\Bigl]
\end{aligned}
\label{thetarho}
\end{equation}
and
\begin{equation}
\begin{aligned}
\theta_{\EH}^v [\bd{g}, \var \bd{g}] &=\f{1}{2}\Bigl(g^{\alpha\beta} \delta \Gamma^v_{\alpha\beta} -g^{\alpha v} \delta\Gamma^\beta_{\alpha\beta}\Bigl) \\
&=\f{1}{2}\Bigl(g^{ab} \delta \Gamma^v_{ab}  +\delta\Gamma^a_{\rho a} \Bigl) \\
&=\f{1}{2}\Bigl( q^{ab} \delta K^{(n)}_{ab} + \var \theta^{(n)} \Bigl).
\end{aligned}
\label{thetav}
\end{equation}
Substituting \eqref{thetarho} and \eqref{thetav} into \eqref{141}, we finally obtain
\begin{equation}
\begin{aligned}
\theta^\mu_\EH \ell_\mu \ \eps_\cN &= -\f{1}{2}\Bigl[2\var(\kappa +\theta^{(n)}) +2\pi_a\var U^a - K^{(\ell)}_{ab} \var q^{ab}  + D_a\var U^a \Bigl]\ \eps_\cN.\\
\end{aligned}
\end{equation}
Concerning the Maxwell pre-symplectic potential, by varying the the Maxwell Lagrangian $\bold{L}_\M = -\f{1}{2}\bd{F}\wedge\star \bd{F}$, we obtain
\begin{equation}
    \var \bold{L}_\M = -\var \bd{A}\wedge \dext\star\bd{F} -\dext(\var\bd{A}\wedge \star \bd{F}),
\end{equation}
where $\bd{F}=\dext\bd{A}$ and the symbol $\star$ identifies the Hodge star operator. The total derivative represents the Maxwell pre-symplectic potential,
\begin{equation}
    \bd{\theta}_\M[\bd{A},\var\bd{A}] = -\var \bd{A}\wedge \star \bd{F}, \qquad i.e. \qquad \bd{\theta}^\cN_\M[\bd{A},\var\bd{A}] = -2\ell_\mu F^{\mu\nu}\var A_\nu\ \eps_\cN.
\end{equation}
In conclusion, the pre-symplectic potential reads
\begin{equation}
\bd{\theta}^\cN_\EM[\bd{g},\bd{A};\var\bd{g},\var\bd{A}] = \bd{\theta}^\cN_\EH[\bd{g};\var\bd{g}] + \bd{\theta}^\cN_\M[\bd{A},\var\bd{A}].
\end{equation}
Let us finally recall that the conserved charges associated with U(1) gauge symmetry $\bd{A}\to \bd{A} + \dext\varepsilon$ are
\begin{equation}
    \cQ^e [\varepsilon] = \int_{\pa\cN} \varepsilon \star \bd{F}, \qquad \cQ^m [\varepsilon] = \int_{\pa\cN} \varepsilon\  \bd{F},
\end{equation}
representing the electric and the magnetic charge, respectively.

\section{Einstein-Cartan formulation}\label{EC}
In this appendix we derive the Noether charges by means the Einstein-Cartan formalism and show where internal symmetry in \eqref{lambda_symm} comes from. Let us first define a null frame $\hat{e}^\mu_{I} = ({\ell}^\mu, {n}^\mu, {m}^\mu, {\bar{m}}^\mu)$, such that ${\ell}\cdot {n}=-1$ and ${m}\cdot {\bar{m}}=1$. We define the frame fields as follows
\begin{equation}
    \hat{e}_0 = \pa_\nu + V\pa_\rho -U^a\pa_a,\qquad
    \hat{e}_1 = \pa_\rho,\qquad
    \hat{e}_i = E_i^a \pa_a
\end{equation}
and the dual frame is
\begin{equation}
\bd{e}^0 = -\dext v, \qquad
\bd{e}^1 =V\dext v - \dext \rho,\qquad
\bd{e}^i = E^i_a (\dext x^a + U^a \dext v),
\end{equation}
where
\begin{equation}
    E_a \dext x^a = \f{1}{\sqrt{2}} \sqrt{ q_{\theta \theta}}\ \dext \theta +\f{1}{\sqrt{2 q_{\theta\theta}}} (q_{\theta\phi} -i\sqrt{q})\ \dext \phi .
\end{equation}
The spin coefficients are defined via the following relation
\begin{equation}
    \omega^{IJ}_\mu \dext x^\mu = e^{I}_\nu\nabla_\mu e^{\nu J}\ \dext  x^\mu,
\end{equation}
and explicitly read
\begin{equation}
\begin{aligned}
\omega^{01}_\mu \dext x^\mu&= \kappa \dext v + \pi_a(U^a\dext v + \dext x^a),\\ 
\omega^{1i}_\mu \dext x^\mu& = E^{ai}\Bigl(\pi_a \dext \rho - K^{(\ell)}_{ab}(\dext x^b + U^b \dext v) +(\pa_a V -V\pi_a) \dext v\Bigl), \\
\omega^{0i}_\mu \dext x^\mu &= -E^{ai} \Bigl(K^{(n)}_{ab}\dext x^b + (\pi_a + U^b K^{(n)}_{ab}) \dext v\Bigl),\\
\omega^{ij}_\mu \dext x^\mu &= E^{[i}_a \pa_\rho E^{j]a}  \dext \rho + \Bigl( E^{[i}_a \pa_v E^{j]a} + E^{a[i} E^{j]}_{\ b} D_a U^b\Bigl)\dext v +\omega^{ij}_{a} \dext x^a.\\
\end{aligned}
\end{equation}
In order to derive the EC charges, we need to define the EC pre-symplectic potential via \eqref{varL=dtheta}. The EC Lagrangian (plus the Holst term) is 
\begin{equation}
    \bold{L}_\ECH= \f{1}{2}\bd{\Sigma}_{IJ}\wedge \bd{R}^{IJ},
\end{equation}
where the curvature tensor is
\begin{equation}
    \bd{R}_{IJ} = \dext\bd{\omega}_{IJ} + \f{1}{2}[\bd{\omega},\bd{\omega}]_{IJ}
\end{equation}
and
\begin{equation}
    \bd{\Sigma}_{IJ}=P_{IJKL}\ \bd{e}^K\wedge \bd{e}^L, \qquad \text{with}\quad P_{IJKL} = \f{1}{2}\epsilon_{IJKL}+\f{1}{\gamma}\eta_{I[K}\eta_{L]J},
\end{equation}
where $\gamma$ is a general parameter. Sometimes $\gamma$ is taken to be the Immirzi parameter, but here we consider it as a general parameter. The symplectic potential is \cite{Freidel:2020svx}
\begin{equation}
    \bd{\theta}_\ECH [\bd{e}, \var\bd{\omega}] = \f{1}{2}\bd{\Sigma}_{IJ}\wedge \var\bd{\omega}^{IJ} = \f{1}{2} P_{IJKL} \bd{e}^K \wedge \bd{e}^L \wedge \var \bd{\omega}^{IJ},
\end{equation}
and does not depend on $\var\bd{e}$. For convenience, we distinguish in the ECH pre-symplectic potential two contributions: the Einstein-Cartan pre-symplectic potential, denoted as follows
\begin{equation}
    \bar{\bd{\theta}}_\EC= \f{1}{4} \epsilon_{IJKL} \bd{e}^K \wedge \bd{e}^L \wedge \var \bd{\omega}^{IJ} ,
\end{equation}
and the dual pre-symplectic potential, which is
\begin{equation}
    \tilde{\bd{\theta}}_\H= \f{1}{2\gamma} \eta_{I[K}\eta_{L]J} \bd{e}^K \wedge \bd{e}^L \wedge \var \bd{\omega}^{IJ}
\end{equation}
and comes from the Holst term. Nonetheless,  we discard the Holst contribution in our analysis and focus solely on the Einstein-Cartan charges.

\subsection*{Symmetries and charges}
The symmetry transformations of the tetrads read as follows
\begin{equation}
    \var_{(\xi,\lambda)} e^I_{\ \mu}= \Lie_\xi  e^I_{\ \mu} - \lambda^I_{\ J} e^J_{\ \mu}.
\end{equation}
The internal gauge transformations have to preserve the structure of the adapted metric, namely 
\begin{equation}
    e^0_{\ \rho}=0, \qquad e^0_{\ a}=0, \qquad e^1_{\ a}=0,
\end{equation}
from which we obtain 
\begin{equation}
\begin{aligned}
    \lambda^{0j} &=- E^{ja}\pa_a \xi^v,\\
    \lambda^{1j} &= E^{ja} (V\pa_a \xi^v -\pa_a \xi^\rho),\\
    \lambda^{10} &= \pa_\rho \xi^\rho,
\end{aligned}
\end{equation}
while $\lambda^{ij}$ remains unfixed. Using the results obtained in \cite{Freidel:2020xyx, Freidel:2020svx}, the Einstein-Cartan charges are defined as follows
\begin{equation}
    \cQ^{\EC}_{(\xi,\lambda)} =\int_\cS \dext^2\sigma \ \sqrt{q} (\iota_\xi \bd{\omega}^{10} +\lambda^{10}).
\end{equation}
Thus, by substituting the leading orders of the spin coefficient $\bd{\omega}^{10}$, we obtain 
\begin{equation}
    \cQ_T^{\EC} = -\int_\cS \dext^2\sigma \sqrt{\mr{q}}\ T\mr{\kappa},  \qquad
    \cQ^{\EC}_Y = -\int_\cS \dext^2\sigma \sqrt{\mr{q}}\  Y^a\mr{\pi}_a, \qquad
    \cQ_W^{\EC} = v\cQ_{T=W}.
\end{equation}
The charge associated with internal gauge transformations yields the so-called internal Lorentz boost charge given by $\lambda^{10}$ and reads
\begin{equation}
    \cQ_\lambda^{\EC} = -\int_\cS \dext^2\sigma \ \sqrt{\mr{q}}\ W.
\end{equation}

\allowdisplaybreaks
\section{List of Christoffel symbols}\label{chrsymb}
The non-vanishing Christoffel symbols are
\begin{align*}
\Gamma^{v}_{vv} &=  \kappa + U^aU^bK^{(n)}_{ab}+ 2 U^a \pi_a \\
\Gamma^{v}_{va} &= \pi_a + U^bK^{(n)}_{ab} \\
\Gamma^{v}_{ab} &= K^{(n)}_{ab} \\
\Gamma^{\rho}_{\rho v} &=  -\kappa - U^b \pi_b\\
\Gamma^{\rho}_{\rho a} &=-\pi_a \\
\Gamma^{\rho}_{vv}&= 2V\kappa +2VU^aU^bK^{(n)}_{ab} + 4VU^a\pi_a -\pa_vV +\f{1}{2}U^aU^b\pa_v q_{ab} -U^a\pa_aV - \f{1}{2}U^aU^bU^c\pa_a q_{bc} \\
&\quad - U^aU_b\pa_a U^b\\
\Gamma^{\rho}_{va} &= U^b K^{(\ell)}_{ab} + U^b V K^{(n)}_{ab} + 2V\pi_{a} -\pa_a V\\
\Gamma^{\rho}_{ab} &= VK^{(n)}_{ab} + K^{(\ell)}_{ab} \\
\Gamma^{a}_{\rho v} &= \pi^{a} + U^b K^a_{(n)b}\\
\Gamma^{a}_{\rho b} &= K^{a}_{(n)b} \\
\Gamma^{a}_{vv} &= - U^a \kappa -U^aU^bU^c K^{(n)}_{bc} - 2U^a U^b\pi_b + q^{ab}\pa_v(q_{bc}U^c)- q^{ab}\pa_b V - \f{1}{2}q^{ab}\pa_b (q_{cd}U^cU^d)\\
\Gamma^{a}_{vb} &= -U^a U^c K^{(n)}_{bc} 
-U^a \pi_{b}+q^{ac}\pa_{[b}(q_{c]d}U^d)  +\f{1}{2}q^{ac}\pa_v q_{cb}\\
\Gamma^{a}_{bc} &= -U^a K^{(n)}_{bc} + \Gamma^a_{bc}[q]  \\
\end{align*}

\bibliography{bib}

\end{document}